\documentclass[aps,floatfix,prd,superscriptaddress,reprint,showpacs,10pt,preprintnumbers,longbibliography]{revtex4-1}
\usepackage[utf8]{inputenc}
\usepackage[pdftex]{graphicx}
\usepackage{float}
\usepackage{amssymb}
\usepackage{amsmath}  
\usepackage{dsfont}
\usepackage{array}
\usepackage{bm}
\usepackage{mathrsfs}
\usepackage{pifont}
\usepackage{multirow}
\usepackage{upgreek}
\usepackage{xcolor}
\usepackage[pdftex,
            pdftitle={Topological vacuum structure of the Schwinger model},
            pdfauthor={Lena Funcke, Karl Jansen, Stefan Kuehn},
            bookmarks,
            colorlinks,
            linkcolor=myblue,
            citecolor=mymagenta,
            menucolor=black,
            urlcolor=myblue,
            plainpages=false,
            pdfpagelabels,
            hypertexnames=false]{hyperref}
\usepackage{verbatim}
\usepackage{slashed}
\usepackage{cleveref}

\definecolor{mymagenta}{RGB}{200, 0, 100}
\definecolor{myblue}{RGB}{45, 48, 146}
\definecolor{mygreen}{RGB}{0, 126, 0}
\definecolor{myorange}{RGB}{255, 136, 19}

\newcommand{\ket}[1]{\mbox{$ | #1 \rangle $}}

\graphicspath{{figures/}}

\begin{document}
\title{Topological vacuum structure of the Schwinger model with matrix product states}
\author{Lena Funcke}
\affiliation{Perimeter Institute for Theoretical Physics, 31 Caroline Street North, Waterloo, ON N2L 2Y5, Canada}
\author{Karl Jansen}
\affiliation{NIC, DESY Zeuthen, Platanenallee 6, 15738 Zeuthen, Germany}
\author{Stefan K{\"u}hn}
\affiliation{Perimeter Institute for Theoretical Physics, 31 Caroline Street North, Waterloo, ON N2L 2Y5, Canada}

\date{\today}
\begin{abstract}

We numerically study the single-flavor Schwinger model with a topological $\theta$-term, which is practically inaccessible by standard lattice Monte Carlo simulations due to the sign problem. By using numerical methods based on tensor networks, especially the one-dimensional matrix product states, we explore the non-trivial $\theta$-dependence of several lattice and continuum quantities in the Hamiltonian formulation. In particular, we compute the ground-state energy, the electric field, the chiral fermion condensate, and the topological vacuum susceptibility for positive, zero, and even negative fermion mass. In the chiral limit, we demonstrate that the continuum model becomes independent of the vacuum angle $\theta$, thus respecting CP invariance, while lattice artifacts still depend on $\theta$. We also confirm that negative masses can be mapped to positive masses by shifting $\theta\rightarrow \theta +\pi$ due to the axial anomaly in the continuum, while lattice artifacts non-trivially distort this mapping. This mass regime is particularly interesting for the (3+1)-dimensional QCD analog of the Schwinger model, the sign problem of which requires the development and testing of new numerical techniques beyond the conventional Monte Carlo approach.

\end{abstract}

\maketitle
\section{Introduction}

QED in 1+1 dimensions, also known as the Schwinger model~\cite{Schwinger1962}, is analytically solvable in the massless-fermion limit and exhibits many properties similar to QCD: confinement, chiral symmetry breaking, a $U(1)_A$ quantum anomaly, and a topologically non-trivial vacuum leading to a $\theta$-term. Therefore, the lattice-regularized version of the Schwinger model has been adopted as a benchmark model for developing and testing new numerical techniques. These comprise algorithms to tackle the sign problem, new ideas for Markov Chain Monte Carlo (MCMC) investigations, and tensor network approaches (see, e.g., Refs.~\cite{Goschl2017,Christian2005,Banuls2018a}, respectively, and references therein).

Already since the seminal work by Coleman and collaborators~\cite{Coleman1975,Coleman1976}, the role of topology in the Schwinger model has been extensively discussed. The non-trivial topological vacuum structure gives rise to a $\theta$-dependent electric background field~\cite{Coleman1975,Coleman1976}, which linearly depends on the fermion mass and gets completely screened in the zero-mass limit~\cite{Adam1997}. Similarly, the ground-state energy density, the chiral fermion condensate, and the topological vacuum susceptibility of the Schwinger model are non-trivially dependent on $\theta$ and the mass parameter. In particular, the topological vacuum susceptibility, which in the QCD analog measures the strength of CP violation of the theory, vanishes in the massless limit.

One well-studied regime in the parameter space of the Schwinger model is the second-order phase transition that occurs at $\theta=\pi$~\cite{Coleman1976} and a fermion mass of $m\approx 0.33g$~\cite{Byrnes2002,Buyens2017}, where $g$ is the dimensionful coupling. A less studied regime is a vanishing or even negative fermion mass, which we consider in the current paper. 

The motivation for looking at this parameter range is twofold. First, it is expected that a negative mass can be trivially mapped to a positive mass by shifting $\theta\rightarrow \theta +\pi$, which is given in the continuum due to the axial quantum anomaly~\cite{Adler1969,Bell1969,Adam1993}. However, as we demonstrate, lattice artifacts distort this mapping, so that negative-mass results only reproduce positive-mass results with $\theta\rightarrow \theta +\pi$ after the continuum extrapolation. The negative-mass scenario is particularly interesting in the many-flavor case, where it can give rise to the CP-violating Dashen phase and pion condensation~\cite{Dashen1970}. 

Second, the zero-mass regime is motivated by the possibility of having a vanishing up-quark mass in the QCD analog, which has been studied for several decades because it provides a potential solution of the long-standing strong CP problem~\cite{Georgi1981,Kaplan1986,Choi1988,Banks1994}. The key point of this proposal is that topological effects can give rise to an effective up-quark mass that does not spoil the solution to the strong CP problem but nevertheless appears in the chiral Lagrangian. There is currently no evidence that this effective mass term is large enough to make the proposal phenomenologically viable~(see Ref.~\cite{Brambilla:2014jmp} for a review), but, if feasible, it would be an elegant solution of the strong CP problem. Both this solution and the alternative axion solution of the strong CP problem~\cite{Peccei1977,Weinberg1977,Wilczek1978} render the theory independent of the vacuum angle $\theta$ in the continuum. However, lattice artifacts still depend on $\theta$, as we show for the single-flavor Schwinger model.

Since all the above questions are of non-perturbative nature, it would be natural to address them through numerical lattice calculations, e.g., by the successful MCMC method. However, the Schwinger model has a sign problem when a topological $\theta$-term is added to the action. This renders the MCMC approach inapplicable, at least when the value of the vacuum angle $\theta$ becomes too large. 

In addition, as discussed above, we are also interested in going to zero and even negative fermion mass. In this case, the lattice Dirac operator $D$ will develop zero modes. This is problematic for standard MCMC methods which use the operator $D^\dagger D$ to have a real and positive action and then integrate out the fermions. The resulting determinant of $D^\dagger D$,
\begin{equation}
{\rm det}(D^\dagger D) \propto 
\int \mathcal{D}\Phi^\dagger\mathcal{D}\Phi\exp\left\{-\Phi^\dagger [D^\dagger D ]^{-1}\Phi\right\},
\label{eq:determinant}
\end{equation}
is estimated stochastically (see, e.g., Ref.~\cite{Gattringer2010}) with appropriately chosen bosonic fields $\Phi^\dagger$ and $\Phi$. Thus, if a zero mode appears in $D^\dagger D$, the integral in Eq.~\eqref{eq:determinant} is ill defined. If we also add a topological $\theta$-term to study the physics questions described above, we encounter a double sign problem, thus ruling out a treatment with MCMC.

A possible way out are tensor network techniques and, since we consider the (1+1)-dimensional Schwinger model, in particular the matrix product state (MPS) approach. Investigations of gauge theories in 1+1 dimensions, especially of the Schwinger model, have progressed substantially over the last years. There have been several works concentrating on the spectrum of the Schwinger model using MPS~\cite{Banuls2013,Banuls2013a,Buyens2013,Kuehn2014,Buyens2014,Buyens2015a,Buyens2015,Banuls2016a,Banuls2016b,Zapp2017,Buyens2017,Banuls2018a}. The model has also been studied at non-zero temperature~\cite{Saito2014,Banuls2015,Saito2015,Buyens2016,Banuls2016,Banuls2018a}, non-zero chemical potential~\cite{Banuls2016a,Banuls2016b,Banuls2016c} and for real-time problems~\cite{Buyens2014,Buyens2016b}. In addition, quantum link models~\cite{Rico2013,Pichler2015}, $\mathds{Z}_n$-QED models~\cite{Ercolessi2018,Magnifico2019,Magnifico:2019ulp}, and non-Abelian gauge models have been explored with the MPS approach~\cite{Kuehn2015,Banuls2017,Sala2018,Sala2018a,Silvi2016,Silvi2019}. Besides MPS, tensor network renormalization techniques~\cite{Shimizu2014,Shimizu2014a,Kawauchi2016} also have been very successfully employed to study properties of gauge theories in 1+1 dimensions and recently even in a simple (2+1)-dimensional gauge theory~\cite{Kuramashi2018}.

An early work on the Schwinger model with a topological term using density matrix renormalization group methods can be found in Ref.~\cite{Byrnes2002} and a more recent one using MPS in Ref.~\cite{Buyens2017}. However, in these papers the main interest has been to explore the phase transition at a value of the vacuum angle $\theta=\pi$. Reference~\cite{Buyens2017} nicely demonstrated the breaking of the CP symmetry happening at a first-order phase transition.

In the present work, we are interested in a different regime of the Schwinger model with a topological $\theta$-term, namely at positive and negative fermion masses close to zero, i.e., far away from the phase transition. We aim to compute the $\theta$-dependence of the ground-state energy density, the electric field, the chiral fermion condensate, and the topological vacuum susceptibility in this regime, in particular for the CP-conserving case of a vanishing mass. This extended regime allows us to examine the full range of validity of the mass-perturbation theory computations in the Schwinger model~\cite{Adam1997}, including its limitations on the lattice. Our work also serves as a proof of concept that the MPS approach works well even for negative masses, which has not been explored before. The examination of this regime is particularly important for the long-term goal of applying such new numerical techniques to the above-mentioned unresolved issues of QCD.

\section{Model and methods}

\subsection{The Schwinger model}

The massive Schwinger model~\cite{Schwinger1962} describes ($1+1$)-dimensional quantum electrodynamics coupled to a single massive Dirac fermion, with the Lagrangian density
\begin{equation}
{\cal{L}}=\bar{\psi} (i \slashed{\partial}- g \slashed{A }-m )\psi-\frac{1}{4}F_{\mu\nu}F^{\mu \nu} + \frac{g\theta}{4\pi}\varepsilon^{\mu\nu}F_{\mu\nu}.
\label{eq:lagrangian}
\end{equation}
Here, $\psi$ denotes the two-component fermionic field with bare mass $m$ and $A_\mu$ is the $U(1)$ gauge field with coupling constant $g$ and field strength $F_{\mu\nu}=\partial_{\mu}A_{\nu}-\partial_{\nu}A_{\mu}$ (where $\mu,\nu=0,1$). The $\theta$-term in Eq.~\eqref{eq:lagrangian} with $\theta\in [0,2\pi]$ is a total derivative and therefore does not affect the classical equations of motion, but it does affect the quantum spectrum. In terms of the dimensionless parameter $m/g$ of the model, both the massless case $m/g=0$ and the free case $m/g\rightarrow\infty$ can be solved analytically. Therefore, the limits of very small or very large masses can be studied within perturbation theory, while the intermediate regime requires a non-perturbative treatment.

The Hamiltonian density of the massive single-flavor Schwinger model in the temporal gauge, $A_0=0$, reads
\begin{equation}
{\cal{H}}=-i\bar{\psi}\gamma^1(\partial_1-igA_1)\psi+m\bar{\psi}{\psi}+\frac{1}{2}\left({\cal{F}}+\frac{g\theta}{2\pi}\right)^2
\label{eq:continuum_hamiltonian}
\end{equation}
plus an irrelevant constant. The electric field, ${\cal{F}}=-\dot{A}^1$, is fixed by the Gau{\ss} constraint, $\partial_1{\cal{F}}=g\bar{\psi}\gamma^0\psi$, up to an integration constant of $g\theta/2\pi$, which corresponds to an electric background field~\cite{Coleman1976}. The $\theta$-parameter can be shifted between the electric field and the fermion mass term by performing an anomalous axial rotation of the fermionic field (see Appendix~\ref{app:theta_mass} for details). The axial quantum anomaly is also the reason why the Hamiltonian in Eq.~\eqref{eq:continuum_hamiltonian} contains a $\theta$-term at all, even though this term can be stripped away on the classical level when the Hamiltonian is formulated in terms of the electric field~\cite{Jackiw:1983nv,Tong2018}.

The dependence of several observables on the constant electric background field $g\theta/2\pi$ was computed in the continuum Schwinger model for the limit $m/g\ll 1$ with mass-perturbation theory~\cite{Adam1997}. The ground-state energy density in units of the coupling was found to be \footnote{Note that the sign of all observables in Eqs.~\eqref{eq:energy_density}--\eqref{eq:chiral_condensate} is reversed with respect to the prediction in Ref.~\cite{Adam1997} due to an opposite sign convention, as pointed out in Ref.~\cite{Adam:1998tw}.}
\begin{align}
   \begin{aligned}
   \frac{{\cal{E}}_0(m,\theta)}{g^2} = -\frac{m\Sigma}{g^2}\cos(\theta) \, -& \,\pi \left(\frac{m\Sigma}{2g^2}\right)^2\\ &\times\left(\mu_0^2{\cal{E}}_+\cos(2\theta) + \mu_0^2{\cal{E}}_-\right),
   \end{aligned}
   \label{eq:energy_density}
\end{align}
where $\Sigma = ge^{\gamma}/(2\pi^{3/2})$, $\gamma$ is the Euler-Mascheroni constant, and $\mu_0^2{\cal{E}}_+= -8.9139$, $\mu_0^2{\cal{E}}_- = 9.7384$ are numerical constants. Note that the topological cosine structure of the ground-state energy density appears analogously in QCD and axion physics, and plays an important role for axion phenomenology (see, e.g., Ref.~\cite{Peccei:2006as}). 

From the energy density \eqref{eq:energy_density}, one can obtain the electric field density, which is (up to a factor) its derivative with respect to $\theta$. In units of the coupling, it is given by
\begin{align}
   \begin{aligned}
      \frac{{\cal{F}}(m,\theta)}{g} =& \, 2\pi\frac{\partial}{\partial \theta} \frac{{\cal{E}}_0(\theta,m)}{g^2} \\ 
      =& \, 2\pi \frac{m\Sigma}{g^2}\sin(\theta) + \pi^2 \left(\frac{m\Sigma}{g^2}\right)^2 \mu_0^2{\cal{E}}_+\sin(2\theta).
   \end{aligned}
   \label{eq:field_density}
\end{align}
Thus, in the presence of massive fermions, the constant electric background field density ${\cal{F}}(\theta)=g\theta /2\pi$ gets partially screened due to vacuum polarization. In the massless case, the field is completely screened, which can be equivalently described by the elimination of the vacuum $\theta$-angle by an axial fermion rotation.

The second derivative of the energy density with respect to $\theta$ corresponds (up to a sign) to the topological vacuum susceptibility. In units of the coupling, it reads
\begin{align}
   \begin{aligned}
      \frac{\chi_\text{top}(m,\theta)}{g} =& -\frac{\partial^2}{\partial\theta^2} \frac{{\cal{E}}_0(\theta,m)}{g^2}\\
      =& \,-\frac{m\Sigma}{g^2}\cos(\theta) - \pi \left(\frac{m\Sigma}{g^2}\right)^2\mu_0^2{\cal{E}}_+\cos(2\theta).
   \end{aligned}
   \label{eq:susceptibility}
\end{align}
This quantity vanishes in the chiral limit where physics becomes $\theta$ independent, similar to the QCD case, where the topological vacuum susceptibility is a measure for CP violation. Note that the susceptibility in Eq.~\eqref{eq:susceptibility} as well as the ground-state energy in Eq.~\eqref{eq:energy_density} and the electric field in Eq.~\eqref{eq:field_density} are invariant under the simultaneous shifts of $\theta\rightarrow\theta+\pi$ and $m\rightarrow -m$. This is because the $\theta$-parameter can be rotated from the electric field into the fermion mass term, as explained above. Thus, a shift by $\Delta\theta=\pi$ gets compensated for by a change of the mass sign, $m\exp(i\Delta\theta)=-m$.

Finally, one can also compute the chiral fermion condensate ${\cal{C}}=\langle\bar{\psi}\psi\rangle$, which is given by the derivative of the energy density with respect to the bare fermion mass,
\begin{align}
\begin{aligned}
 \frac{{\cal{C}}(m,\theta)}{g} =& \, g \frac{\partial}{\partial m}\frac{{\cal{E}}_0(m,\theta)}{g^2}\\
  =& -\frac{\Sigma}{g}\cos(\theta) - \frac{\pi m}{2g}\left(\frac{\Sigma}{g}\right)^2\\ &\hspace{6.2em}\times\left(\mu_0^2{\cal{E}}_+\cos(2\theta) + \mu_0^2{\cal{E}}_-\right),
   \end{aligned}
\label{eq:chiral_condensate}
\end{align}
and which is independent of the fermion mass in first order. The condensate transforms similarly to the fermion mass under an axial transformation, therefore the above-mentioned shift of $\theta\rightarrow\theta+\pi$ induces $m\rightarrow -m$ and ${\cal{C}}(m,\theta)\rightarrow -{\cal{C}}(-m,\theta+\pi)$, as can be seen in Eq.~\eqref{eq:chiral_condensate}. In the massless limit, the condensate still seems $\theta$ dependent at first sight, but this angular parameter becomes unphysical as it can be rotated away by said axial rotation. Equivalently, for $m=0$ the phase of the condensate can be absorbed by a shift in the Schwinger boson field and becomes unobservable. Also note that the condensate itself it not a physical quantity and only enters observable quantities when multiplied by the fermion mass. Thus, the model's $\theta$-dependence still vanishes for $m=0$.

\subsection{Lattice formulation}

Our goal is to numerically compute the $\theta$-dependence of the quantities in Eqs.~\eqref{eq:energy_density}--\eqref{eq:chiral_condensate} using the MPS approach, for positive, zero, and negative fermion masses. For our numerical simulations with MPS, we use a lattice formulation of the Schwinger Hamiltonian. To distinguish lattice from continuum quantities in our equations, we denote lattice quantities with roman letters as opposed to the calligraphic letters for continuum quantities.

A possible discretization of the continuum Schwinger Hamiltonian in Eq.~\eqref{eq:continuum_hamiltonian} on a lattice with spacing $a$ is given by the Kogut-Susskind Hamiltonian~\cite{Kogut1975}
\begin{align}
   \begin{aligned}
      H = &-\frac{i}{2a}\sum_{n}\left(\phi^\dagger_{n}e^{i\vartheta_n}\phi_{n+1}-\mathrm{h.c.}\right) \\ &+m\sum_{n}(-1)^n \phi^\dagger_{n}\phi_{n} + \frac{ag^2}{2}\sum_{n} F_n^2.
   \end{aligned}
   \label{eq:hamiltonian}
\end{align}
In the expression above, $\phi_{n}$ is a single-component fermionic field describing a fermion on site $n$, $m$ is the bare fermion mass, and $g$ is the coupling constant. The operators $F_n$ and $\vartheta_n$ act on the gauge links in between the fermions, and $F_n$ gives the quantized electric flux on link $n$. They fulfill the commutation relation $[\vartheta_n,F_k]=i\delta_{n,k}$ and hence $e^{i\vartheta_n}$ acts as rising operator for the electric flux. The angle $\vartheta_n$ is restricted to $[0,2\pi]$ since we use a compact formulation. The Gau{\ss} constraint on the physical states translates to
\begin{align}
   F_n - F_{n-1} = Q_n\quad\quad\forall n
   \label{eq:gauss_law_lattice}
\end{align}
on the lattice, where $Q_n = \phi^\dagger_n\phi_n - \bigl(1-(-1)^n\bigr)/2$ is the staggered fermionic charge.

We choose to work with open boundary conditions, for which we can use Eq.~\eqref{eq:gauss_law_lattice} to integrate out the gauge fields~\cite{Hamer1997,Banuls2013}. After a residual gauge transformation, the dimensionless lattice Hamiltonian reads
\begin{align}
   \begin{aligned}
   \begin{split}
      W =& \frac{2}{ag^2}H\\ =& -ix\sum_{n=0}^{N-2}\left(\phi^\dagger_n\phi_{n+1}-\mathrm{h.c.}\right) \\ &+ \mu \sum_{n=0}^{N-1}(-1)^n \phi^\dagger_{n}\phi_{n} + \sum_{n=0}^{N-2} \left(\sum_{k=0}^nQ_k+\frac{\theta}{2\pi}\right)^2,
      \end{split}
   \end{aligned}
   \label{eq:hamiltonian_dimless}
\end{align}
where we have defined the dimensionless constants $x\equiv 1/(ag)^2$ and $\mu \equiv \sqrt{x}m/g$. As in the continuum case, the dimensionless integration constant $\theta/2\pi$ corresponds to a constant electric background field. Hence, we see that the model only has three independent parameters: the lattice spacing and the bare fermion mass, both in units of the coupling, and the vacuum angle $\theta$. 

We would like to make contact to the continuum prediction with the results extracted from the dimensionless lattice Hamiltonian. To this end, let $E_0(m,\theta)$ be the ground-state energy of the dimensionless Hamiltonian $W$ from Eq.~\eqref{eq:hamiltonian_dimless}, which is related to the dimensionful energy density ${\cal{E}}_0(m,\theta)$ from Eq.~\eqref{eq:energy_density} by
\begin{align}
   \begin{aligned}
      E_0(m,\theta) = \frac{2}{ag^2}{\cal{E}}_0(m,\theta)L = \frac{2\sqrt{x}}{g}{\cal{E}}_0(m,\theta)L.
   \end{aligned}
   \label{eq:E_conversion}
\end{align}
Thus, starting from ${\cal{E}}_0(m,\theta)$, we find
\begin{align}
   {\cal{E}}_0(m,\theta) = g^2\frac{E_0(m,\theta)}{2N},
\end{align}
where we have used that the volume $L$ in units of the coupling is given by $Lg = N/\sqrt{x}$. Consequently, the lattice quantity that should correspond to the continuum energy density~\eqref{eq:energy_density} is $E_0/2N$. Note, however, that $E_0/2N$ is UV divergent~\cite{Adam:1998tw}. In order to obtain a UV-finite quantity, we can simply subtract the result for a fixed value of $\theta = \theta_0$ and look at $\Delta E_0(m,\theta)/2N = [E_0(m,\theta) - E_0(m,\theta_0)]/2N$. For small enough bare fermion mass, we expect that toward the continuum limit this UV-finite lattice ground-state energy density becomes approximately equal to the perturbative continuum prediction $\Delta{\cal{E}}_0(m,\theta)$~\footnote{Note that the quantity \eqref{eq:UV-finite_energy} corresponds to the string tension in Ref.~\cite{Buyens2017}.},
\begin{align}
      \frac{\Delta E_0(m,\theta) }{2N} \approx & \frac{\Delta{\cal{E}}_0(m,\theta) }{g^2} = \frac{{\cal{E}}_0(m,\theta) - {\cal{E}}_0(m,\theta_0)}{g^2}.
   \label{eq:UV-finite_energy}
\end{align}
After extrapolating our numerical lattice data for the UV-finite ground-state energy density $\Delta E_0(m,\theta)$ to the continuum, we can numerically compute the derivatives to obtain the continuum electric field density ${\cal{F}}(m,\theta)$ and the continuum topological vacuum susceptibility $\chi(m,\theta)$. The electric field density and the topological vacuum susceptibility are already UV finite without subtracting their values at $\theta_0$, which is why we denote them as $F$ and $\chi$ instead of $\Delta F$ and $\Delta \chi$, respectively.

Alternatively, we can also directly measure the electric field per unit volume with MPS and numerically compute its derivative to get the continuum topological vacuum susceptibility. As $(F_n+\theta/2\pi)g$ approaches the electric field ${\cal{F}}(x)$ in the continuum limit, we expect $F_n+\theta/2\pi$ to follow Eq.~\eqref{eq:field_density} for small fermion masses toward the continuum. Notice that we are using a staggered formulation; thus, in order to compensate for staggering effects, we average $F_n$ over the system and look at the quantity
\begin{align}
F_\text{av} = \sum_{n=0}^{N-2} \frac{F_n+\theta/2\pi}{N-1}.
\end{align}
In addition, we also have direct access to the chiral condensate, which in our staggered formulation translates to $C=\sqrt{x}\sum_{n=0}^{N-1}(-1)^n\phi^\dagger_n\phi_n/N$. For nonvanishing bare fermion mass, this quantity is UV divergent~\cite{Banuls2013a,Buyens2014,Banuls2016}. Thus, we again subtract the value for $\theta_0$, $\Delta C(m,\theta) = C(m,\theta) - C(m,\theta_0)$, and expect to find for small fermion masses toward the continuum limit
\begin{align}
      \Delta C(m,\theta)&\approx \frac{\Delta{\cal{C}}(m,\theta)}{g} = \frac{{\cal{C}}(m,\theta) - {\cal{C}}(m,\theta_0)}{g}.
   \label{eq:UV-finite_condensate}
\end{align}

\subsection{Matrix product states}

In order to obtain the ground state of the Hamiltonian in Eq.~\eqref{eq:hamiltonian_dimless}, we use the MPS ansatz. For a system with $N$ sites and open boundary conditions, it reads
\begin{align}
   \ket{\psi} = \sum_{i_1,i_2,\dots,i_N}^d A^{i_1}_1A^{i_2}_2\cdots A^{i_N}_N\ket{i_1}\otimes\ket{i_2}\otimes\dots\otimes\ket{i_N},
\end{align}
where $A^{i_k}_k$ are complex matrices of size $D\times D$ for $1<k<N$ and $A^{i_1}_1$ ($A^{i_N}_N$) is a $D$-dimensional row (column) vector. The size $D$ of the matrices, called the bond dimension of the MPS, determines the number of variational parameters in the ansatz and limits the amount of entanglement that can be present in the state (see Refs.~\cite{Verstraete2008,Schollwoeck2011,Orus2014a} for detailed reviews).

Given a Hamiltonian, the MPS approximation for the ground state can be found in a standard manner by iteratively updating the tensors $A^{i_k}_k$ one after another while keeping the others fixed~\cite{Verstraete2004}. In each step, the optimal tensor is determined by finding the ground state of an effective Hamiltonian describing the interaction of the site with its environment. The ground-state wave function is obtained by repeating the updating procedure starting from the left boundary and sweeping back and forth until the relative change of the energy is below a certain tolerance $\eta$. After obtaining the MPS for the ground state, we can measure all kinds of (local) quantities such as the electric field and the chiral condensate.

For convenience in the simulations, we choose to translate the fermionic degrees of freedom in Eq.~\eqref{eq:hamiltonian_dimless} to spins using a Jordan-Wigner transformation~\cite{Banuls2013}. Although tensor networks and in particular MPS can deal with fermionic degrees of freedom with essentially no additional cost in the algorithm~\cite{Corboz2009,Pineda2010,Kraus2010}, this allows us to avoid dealing with anticommuting fermionic operators. In spin language, Eq.~\eqref{eq:hamiltonian_dimless} reads
\begin{align}
   \begin{aligned}
      W =\, &  x\sum_{n=0}^{N-2}\left(\sigma^+_n\sigma_{n+1}^- +\mathrm{h.c.}\right) + \frac{\mu}{2} \sum_{n=0}^{N-1}(-1)^n (\mathds{1} +\sigma_n^z)\\
      & + \sum_{n=0}^{N-2} \left(\sum_{k=0}^nQ_k+\frac{\theta}{2\pi}\right)^2,
   \end{aligned}  
   \label{eq:spin_hamiltonian} 
\end{align}
where $Q_n = \left(\sigma^z_n +(-1)^n\right)/2$ is the staggered charge and $\sigma^\pm = (\sigma^x \pm i\sigma^y)$ and $\sigma^z$ are the usual Pauli matrices.

Although the Hamiltonian in Eq.~\eqref{eq:spin_hamiltonian} is nonlocal, we expect MPS to be a suitable ansatz to describe its low-energy spectrum. For one, the original Hamiltonian~\eqref{eq:hamiltonian} is local, and for bare fermion masses smaller than $(m/g)_c\approx 0.33$ the model is gapped with a unique ground state. Moreover, as shown in Ref.~\cite{Buyens2017}, its low-energy states are characterized by small values of the electric field. Hence, the gauge links can be effectively considered as finite dimensional and the ground state can be efficiently described by a MPS~\cite{Hastings2007}. Integrating the gauge degrees of freedom out can be seen as projecting the ground state of the original Hamiltonian locally on the gauge invariant subspace. Such a projection increases the bond dimension at maximum by a factor on the order of the effective dimension of the gauge links. As a result, the ground state of Hamiltonian~\eqref{eq:spin_hamiltonian} is expected to be well described by a MPS with small bond dimension, too.

\section{Results}

We examine the $\theta$-dependence of the ground-state energy density, the electric field, the chiral condensate, and the topological vacuum susceptibility for a wide range of fermion masses, $m/g\in[-0.07,0.21]$, and lattice spacings corresponding to $x=1/(ag)^2\in[80,160]$. In order to systematically probe for finite-volume effects and to be able to extrapolate our results to the thermodynamic limit, we explore for each combination of $(\theta,m/g,x)$ a large range of system sizes corresponding to volumes $N/\sqrt{x}\in[4.5,45]$. In addition, we have another truncation effect due to the finite bond dimension present in our numerical simulations. This error can be controlled by repeating the calculation for every set of $(\theta,m/g,x,N)$ for a range of $D\in[20,140]$ and extrapolating to the limit $D\to\infty$ (details about the extrapolation procedure can be found in Appendix~\ref{app:extrapolation}). In all our simulations, we stop as soon as the relative change in the ground-state energy is below $\eta=10^{-10}$. Moreover, we focus on half a period of $\theta\in[0,\pi]$, since all observable quantities are predicted to be (point) symmetric around $\theta=\pi$ (see Eqs.\ \eqref{eq:energy_density}--\eqref{eq:chiral_condensate} and exemplary data of a full period of $\theta\in[0,2\pi]$ in Appendix~\ref{app:fullperiod}).

\subsection{Ground-state energy}

\begin{figure}
   \centering
   \includegraphics[width=0.48\textwidth]{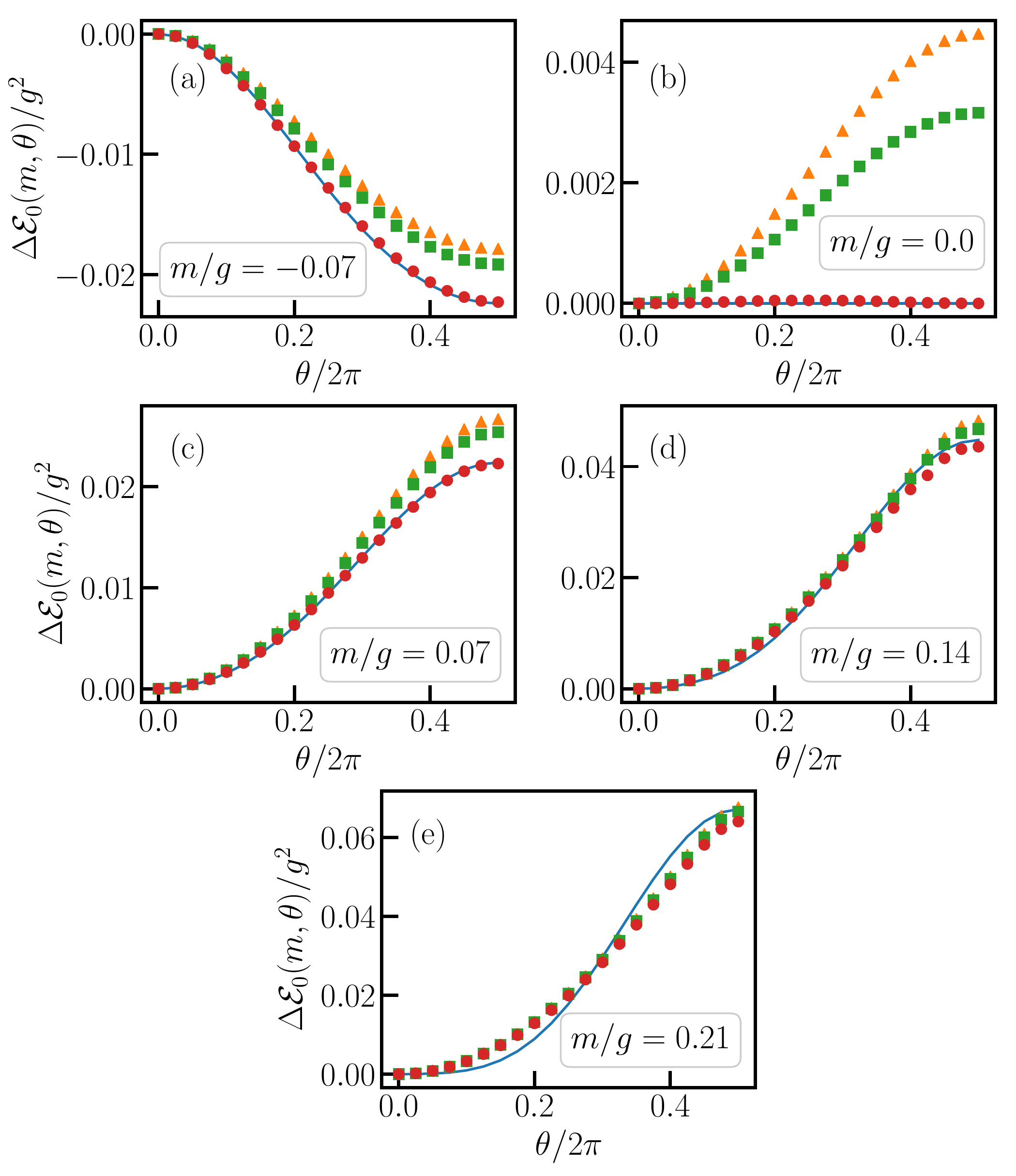}
   \caption{UV-finite ground-state energy density as a function of the angle $\theta$ for (a) $m/g=-0.07$, (b) $m/g=0.0$, (c) $m/g=0.07$, (d) $m/g=0.14$, and (e) $m/g=0.21$. The orange triangles (green squares) correspond to finite-lattice data with $x=80$ ($x=160$).  The red dots represent the result obtained after extrapolating our finite-lattice data to the continuum which can be compared to the perturbative prediction from Eq.~\eqref{eq:UV-finite_energy} (blue solid line). In all cases the error bars are smaller than the markers.}
   \label{fig:energy_vs_theta}
\end{figure}

Our results for the UV-finite ground-state energy density are shown in Fig.~\ref{fig:energy_vs_theta}, after subtracting the value for $\theta_0=0$ and extrapolating to the limit $N\to\infty$. The figure contains our data for the largest and smallest lattice spacing as well as the continuum extrapolation. Note that the result for $\theta_0=0$, which is subtracted in $\Delta{\cal{E}}_0(m,\theta)$, is smaller than the results for $\theta>\theta_0$. This is why the UV-finite ground-state energy density in Fig.~\ref{fig:energy_vs_theta} is positive for $m>0$, while the UV-infinite expression in Eq.~\eqref{eq:energy_density} is negative.

In general, we observe that we can reliably extrapolate to the thermodynamic limit and control our errors due to the finite bond dimension and system size (see Appendix~\ref{app:extrapolation} for details). Independent of the fermion mass, we see that finite-lattice effects are more pronounced around the extremal values of the energy density at $\theta=\pi$. For small masses, our data exhibit larger relative changes when going to smaller lattice spacings, whereas there is hardly any shift for the two largest masses $m/g=0.14$ and $0.21$, especially for small $\theta$. 

With our finite-lattice data, we can extrapolate to the limit $ag\to 0$ and estimate the continuum values. As Fig.~\ref{fig:energy_vs_theta} reveals, the error bars resulting from this extrapolation are negligible and we can obtain precise estimates for the continuum limit. Comparing our results to the prediction from mass-perturbation theory in Eq.~\eqref{eq:UV-finite_energy}, we observe excellent agreement for $m/g\leq 0.07$ (see Figs.~\ref{fig:energy_vs_theta}(a)--(c)). For the largest two masses, perturbation theory eventually breaks down and fails to describe our data, as expected. Even though $m/g=0.21$ is still much smaller than unity,  the perturbative results become less reliable than our numerical computations in this regime. This is illustrated by comparing the perturbative prediction of the critical mass of the phase transition, $(m/g)_c\approx 0.18$ (based on Ref.~\cite{Adam:1998tx}), with the non-perturbative value $(m/g)_c\approx 0.18\ll 0.33$. Although the perturbative calculations in Ref.~\cite{Adam1997} are still qualitatively correct and approximately follow our numerical data even in the large-mass parameter regime (see Figs.~\ref{fig:energy_vs_theta}(d) and \ref{fig:energy_vs_theta}(e)), we can only obtain quantitatively precise results with numerical techniques beyond perturbation theory.

In particular, it is interesting to look at our continuum data for vanishing bare fermion mass. As predicted by the perturbative result in Eq.~\eqref{eq:UV-finite_energy}, we indeed observe that the energy density becomes independent of $\theta$ once the extrapolation to the continuum is performed. This elimination of the $\theta$-parameter in the chiral limit is only given in the continuum and does not apply to finite lattice spacings, as one can see in Fig.~\ref{fig:energy_vs_theta}(b).

Moreover, it is instructive to compare our results for $m/g=-0.07$ and $0.07$. In the continuum, a negative bare fermion mass can be mapped to the same positive mass value by shifting the $\theta$-angle by $\pi$ (see Eq.~\eqref{eq:UV-finite_energy}). Hence, a negative mass yields the same ground-state energy for $\theta\in[0,\pi]$ as the corresponding positive mass for $\theta\in[\pi,2\pi]$. This can be seen when comparing Figs.~\ref{fig:energy_vs_theta}(a) and \ref{fig:energy_vs_theta}(c), keeping in mind that the mapping requires not only the shift $\theta\rightarrow\theta +\pi$ but also the shift $\theta_0\rightarrow\theta_0 +\pi$, i.e., the UV-finite quantity is obtained by subtracting the value at $\theta_0=\pi$ instead of $\theta_0=0$. 

\begin{figure}
   \centering
   \includegraphics[width=0.48\textwidth]{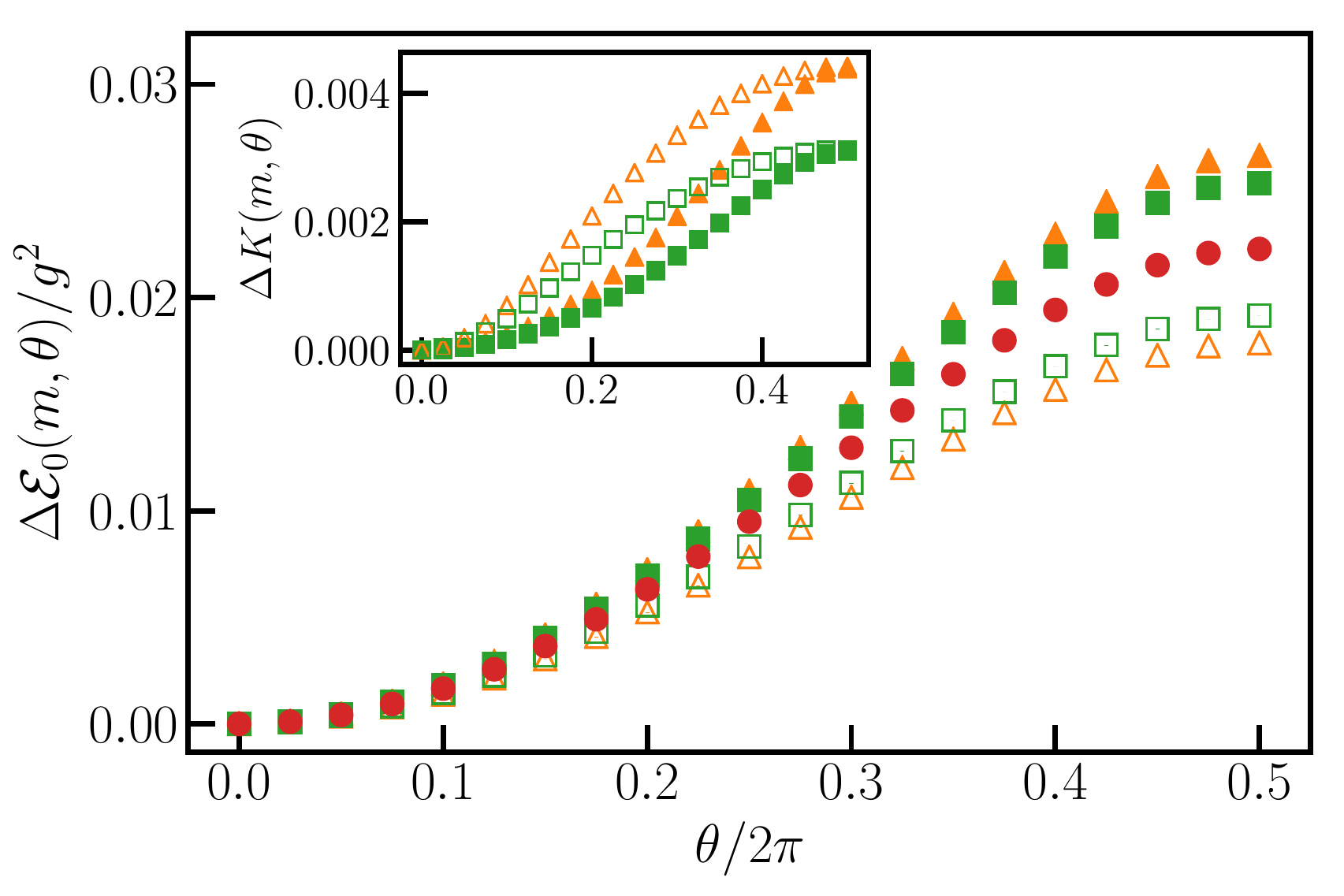}
   \caption{Comparison between the finite-lattice and continuum data of the ground-state energy density from Fig.~\ref{fig:energy_vs_theta} for $m/g=0.07$ (filled markers) and for $m/g=-0.07$ after reflection with respect to the $x$ axis (open markers). The orange triangles (green squares) correspond to $x=80$ ($x=160$), the red dots to the continuum limit. Inset: Absolute value of deviations between the lattice and continuum data (see Eq.~\eqref{eq:deviation}) for $m/g=0.07$ (filled markers) and $m/g=-0.07$ (open markers) for $x=80$ (orange triangles) and $x=160$ (green squares).}
   \label{fig:cmp_lattice_effects}
\end{figure}

Equivalently, the continuum data for $m<0$ can be mapped to the continuum data for $m>0$ by reflecting the former with respect to the $x$ axis. Even though this mapping works well in the continuum, Fig.~\ref{fig:cmp_lattice_effects} reveals that it is distorted for our finite-lattice data. This is expected due to the difficulty to establish chiral symmetry and the corresponding axial anomaly on the lattice~\footnote{For a discussion of chiral invariance on the lattice, see e.g.\ the overview in Ref.~\cite{Niedermayer:1998bi}.}.

The artifacts from the finite lattice spacing enter with opposite sign, breaking the reflection symmetry for negative and positive masses. Moreover, for a fixed value of the lattice spacing, the deviations from the continuum result not only differ in sign, but also their absolute value
\begin{equation}
\Delta K(m,\theta) = \left|\frac{\Delta E_0(m,\theta)}{2N} - \frac{\Delta \mathcal{E}_{0}(m,\theta)}{g^2}\right|
\label{eq:deviation}
\end{equation}
differs in magnitude, especially for intermediate values of $\theta$ (see inset of Fig.~\ref{fig:cmp_lattice_effects}). While the deviations for positive and negative bare fermion masses are comparable for $\theta\approx 0$ and $\theta\approx \pi$, the data for positive $m/g$ have smaller lattice effects in the intermediate regime, even though the order of magnitude is the same in both cases. 

Looking at Figs.~\ref{fig:energy_vs_theta}(a) and \ref{fig:energy_vs_theta}(c), we see that these differences disappear when extrapolating to the limit of vanishing lattice spacing. Thus, the reflection symmetry is restored and our continuum data is in excellent agreement with the perturbative prediction for small masses.

\subsection{Electric field}

\begin{figure}
   \centering
   \includegraphics[width=0.48\textwidth]{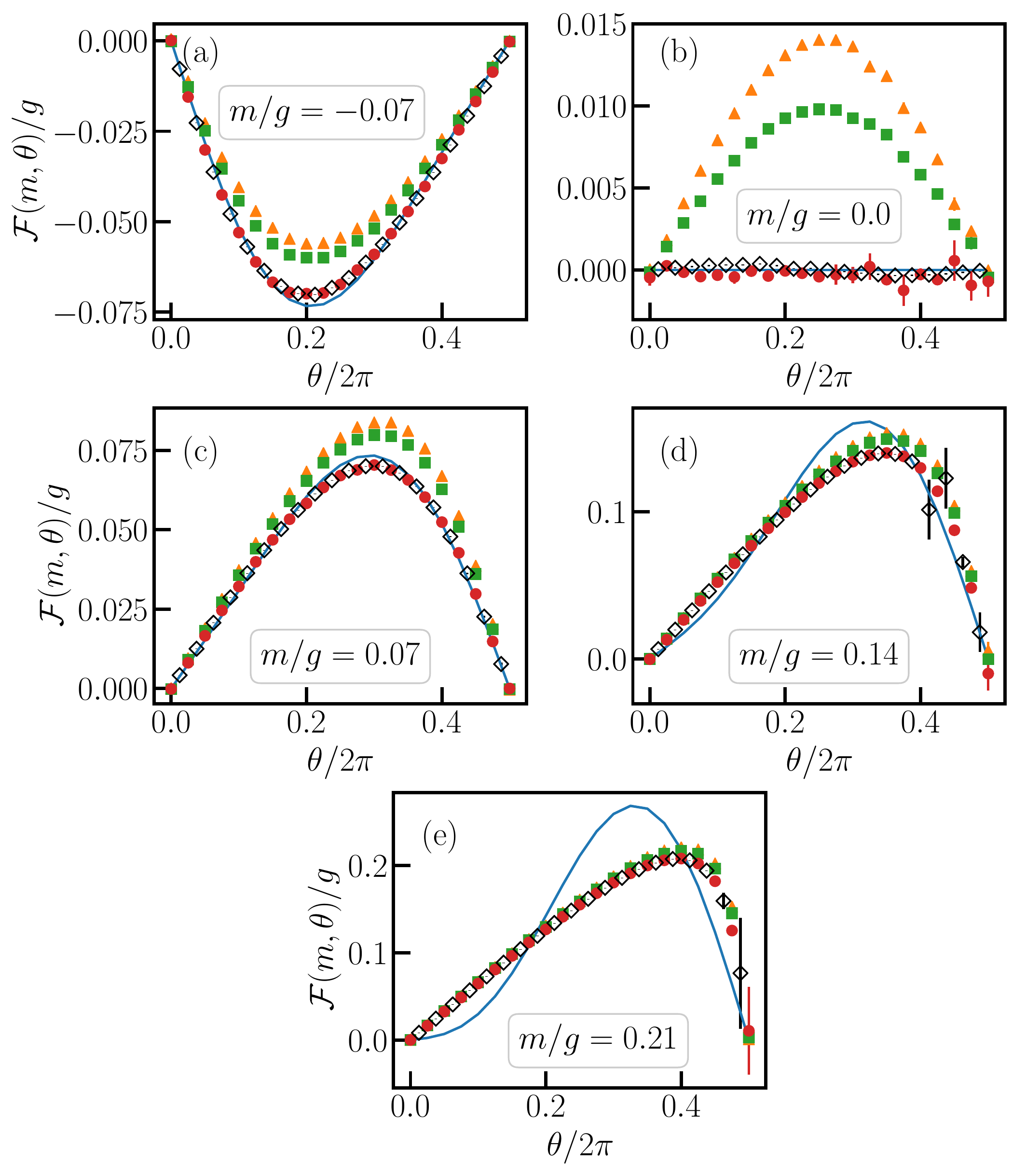}
   \caption{Electric field density as a function of the angle $\theta$ for (a) $m/g=-0.07$, (b) $m/g=0.0$, (c) $m/g=0.07$, (d) $m/g=0.14$, and (e) $m/g=0.21$. The orange triangles (green squares) correspond to finite-lattice data with $x=80$ ($x=160$). The red dots represent the result obtained after extrapolating our finite-lattice data to the continuum which can be compared to the perturbative prediction from Eq.~\eqref{eq:field_density} (blue solid line). In addition, we also show the results obtained by numerically computing the derivative of our continuum estimate for the UV-finite ground-state energy density (open black diamonds).}
   \label{fig:field_vs_theta}
\end{figure}

With our MPS approach, we can directly measure the electric field in the ground state. Performing the same extrapolation procedure as for the energy density, we obtain the data shown in Fig.~\ref{fig:field_vs_theta}. Again, the errors resulting from the extrapolation in system size and bond dimension are negligible, and finite-lattice effects are more pronounced for smaller values of $m/g$. The lattice effects become stronger for $\theta = \pi/4$ due to the sine dependence of the electric field (see Eq.~\eqref{eq:field_density}), in contrast to the ground-state energy in Fig.~\ref{fig:energy_vs_theta} whose cosine-dependent lattice effects become more distinct for $\theta=\pi$ (see Eq.~\eqref{eq:energy_density}).

Just as before, we can estimate the continuum value of the electric field by extrapolating our finite-lattice results to the limit of vanishing lattice spacing. In general, we find that we can reliably estimate the continuum limit, while the errors are slightly larger compared to the energy density. In particular, for the largest two masses we observe enhanced error bars for a $\theta$-angle of $\pi$, which is likely caused by the fact that the continuum extrapolation becomes increasingly challenging as we approach the phase transition at $(m/g)_c\approx 0.33$ and $\theta=\pi$.

When comparing our continuum results for the electric field to the predictions from mass-perturbation theory (see Eq.~\eqref{eq:field_density}), we observe a similar picture as for the ground-state energy density. For bare fermion masses $m/g\leq 0.07$ (see Figs.~\ref{fig:field_vs_theta}(a)--(c)), the perturbative prediction is in excellent agreement with our continuum results. In particular, for $m/g=0$ our data are compatible with zero, independent of $\theta$. Hence, our results confirm that for vanishing mass, the background field gets screened completely and the total electric field vanishes. Moreover, our data for $m/g=-0.07$ and $m/g=0.07$ can be mapped in the continuum by reflection with respect to the $x$ axis, while finite-lattice artifacts distort this mapping with opposite sign. Moving on to larger fermion masses, perturbation theory eventually breaks down and Eq.~\eqref{eq:field_density} does not reproduce the behavior of our data for $m/g\geq 0.14$. In contrast to the energy density, the perturbative result does not even qualitatively describe our numerical data for the largest mass, $m/g =0.21$. As pointed out before, this mass regime cannot be accurately captured by perturbation theory, which manifests itself in predicting a wrong critical mass of the phase transition.

As a cross-check, we can also obtain data for the electric field density by numerically computing the derivative of our results for the continuum energy density, which we show in Fig.~\ref{fig:field_vs_theta} for comparison. In general, the values for the electric field obtained in this way are in good agreement with those from extrapolating the direct measurement of the electric field to the continuum. Although numerically computing the derivative enhances the errors by a factor proportional to 
\begin{figure}
   \centering
   \includegraphics[width=0.48\textwidth]{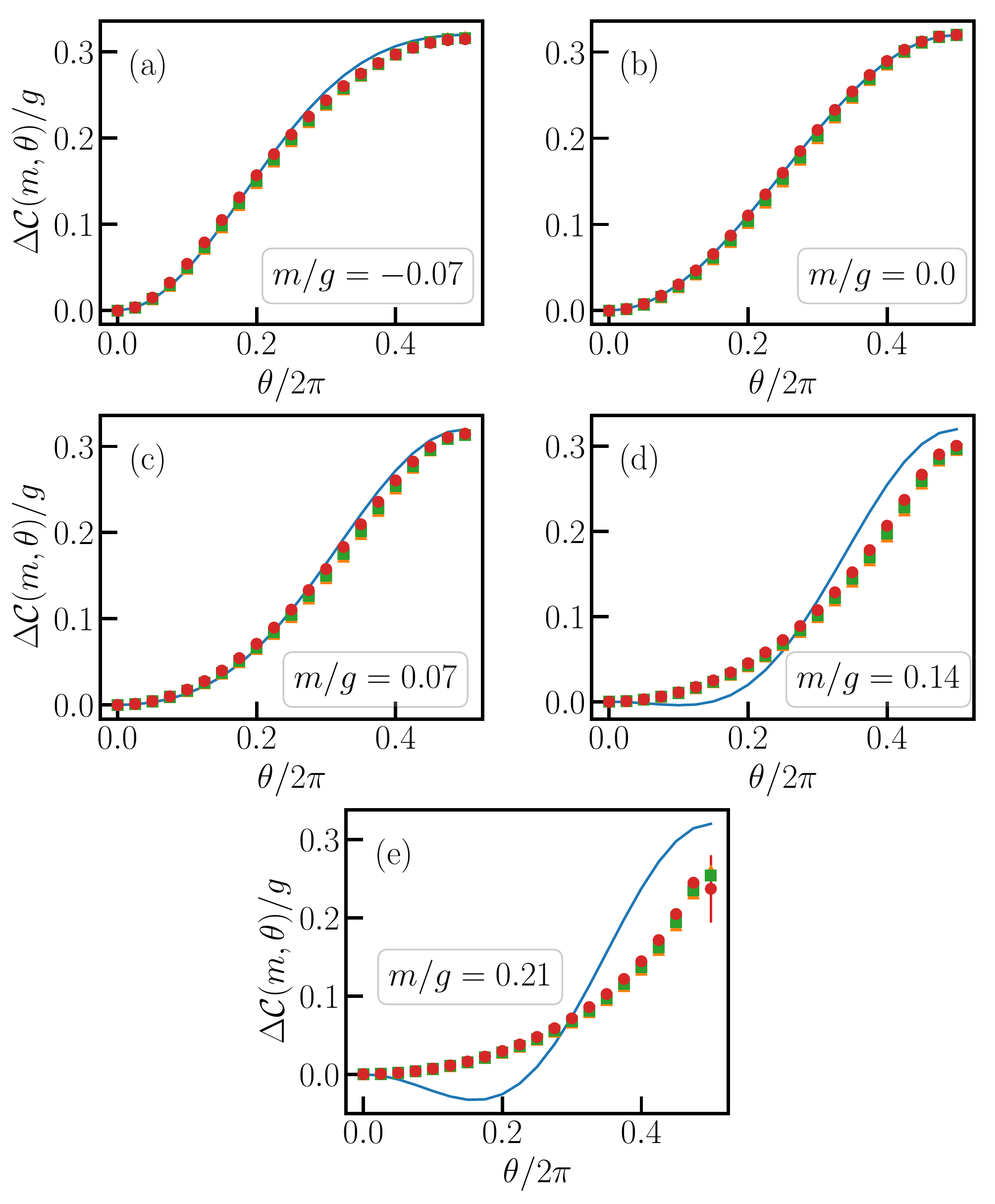}
   \caption{UV-finite chiral condensate as a function of the angle $\theta$ for (a) $m/g=-0.07$, (b) $m/g=0.0$, (c) $m/g=0.07$, (d) $m/g=0.14$, and (e) $m/g=0.21$. The orange triangles (green squares) correspond to finite-lattice data with $x=80$ ($x=160$). The red dots represent the result obtained after extrapolating our finite-lattice data to the continuum which can be compared to the perturbative prediction from Eq.~\eqref{eq:UV-finite_condensate} (blue solid line). Note that the sign of the data for positive and negative masses does not change since the chiral condensate is mass independent at leading order.}
   \label{fig:condensate_vs_theta}
\end{figure}
$1/\Delta \theta$, the results from our data for the energy density are typically more precise than the extrapolated electric field values. The reason for this is that the electric field is in general more sensitive to finite-volume and finite-lattice effects (see Appendix~\ref{app:extrapolation}), resulting in larger error bars in the extrapolated values. For large fermion masses, the errors increase in both cases around $\theta \approx \pi$, thus indicating that we get closer to the critical value $(m/g)_c$.

\subsection{Chiral condensate}

The MPS approach also gives us access to the chiral condensate in the ground state. Performing the same extrapolation procedure as for the ground-state energy density and the electric field, we obtain the results in Fig.~\ref{fig:condensate_vs_theta}, where we have again subtracted the value for $\theta_0 = 0$. As before, the result for $\theta_0=0$, which is subtracted in $\Delta{\cal{C}}_0(m,\theta)$, is smaller than the results for $\theta>\theta_0$. This is why the UV-finite chiral condensate in Fig.~\ref{fig:condensate_vs_theta} is positive for $m>0$, while the UV-infinite expression in Eq.~\eqref{eq:chiral_condensate} is negative.

Compared to the ground-state energy density and the electric field, the chiral condensate is less susceptible to finite-lattice effects and there is hardly any difference between results for our coarsest and finest lattice spacing. As a result, the error bars from extrapolating our finite-lattice data to the limit $ag\to 0$ are essentially negligible.

Notice that the chiral condensate corresponds to the derivative of the energy density with respect to the bare fermion mass (see Eq.~\eqref{eq:chiral_condensate}). Thus, in leading order, mass-perturbation theory predicts a behavior independent of the parameter $m/g$. Looking at our results for small masses in Figs.~\ref{fig:condensate_vs_theta}(a)--(c), we see that this indeed is the case, and Eq.~\eqref{eq:UV-finite_condensate} is in excellent agreement with our data. 

At next order in perturbation theory, we expect finite-mass effects to become relevant, which becomes particularly interesting when comparing our data for $m/g=-0.07$ and $0.07$. Since the chiral condensate transforms similarly to the fermion mass under an axial rotation, a shift of $\theta\rightarrow\theta+\pi$ does not only induce $m\rightarrow -m$ but also ${\cal{C}}(m,\theta)\rightarrow -{\cal{C}}(-m,\theta+\pi)$ (see Eq.~\eqref{eq:UV-finite_condensate}). Hence, changing the sign of the condensate for $m<0$ in the range $\theta\in[0,\pi]$ reproduces the corresponding positive condensate for $m>0$ in the range $\theta\in[\pi,2\pi]$. This can be seen in Figs.~\ref{fig:condensate_vs_theta}(a) and \ref{fig:condensate_vs_theta}(c), keeping in mind that the shifted UV-finite quantity is obtained by subtracting the value at $\theta_0=\pi$ instead of $\theta_0=0$.

In the massless limit, our data for the chiral condensate are still $\theta$ dependent (see Fig.~\ref{fig:condensate_vs_theta}(b)), but $\theta$ becomes an unphysical parameter as it can be rotated away by the above-mentioned axial rotation (see Appendix~\ref{app:theta_mass}). For larger values of $m/g$, the data only change moderately (see Figs.~\ref{fig:condensate_vs_theta}(d) and \ref{fig:condensate_vs_theta}(e)), whereas perturbation theory erroneously predicts new qualitative features that do not occur in our numerical data, such as a dip around $\theta/2\pi \approx 0.19$. Thus, similar to the electric field, the prediction by mass-perturbation theory for the chiral condensate in Eq.~\eqref{eq:chiral_condensate} breaks down for large masses $m/g\geq 0.14$ as we approach the phase transition.

\subsection{Topological vacuum susceptibility}

\begin{figure}
   \centering
   \includegraphics[width=0.48\textwidth]{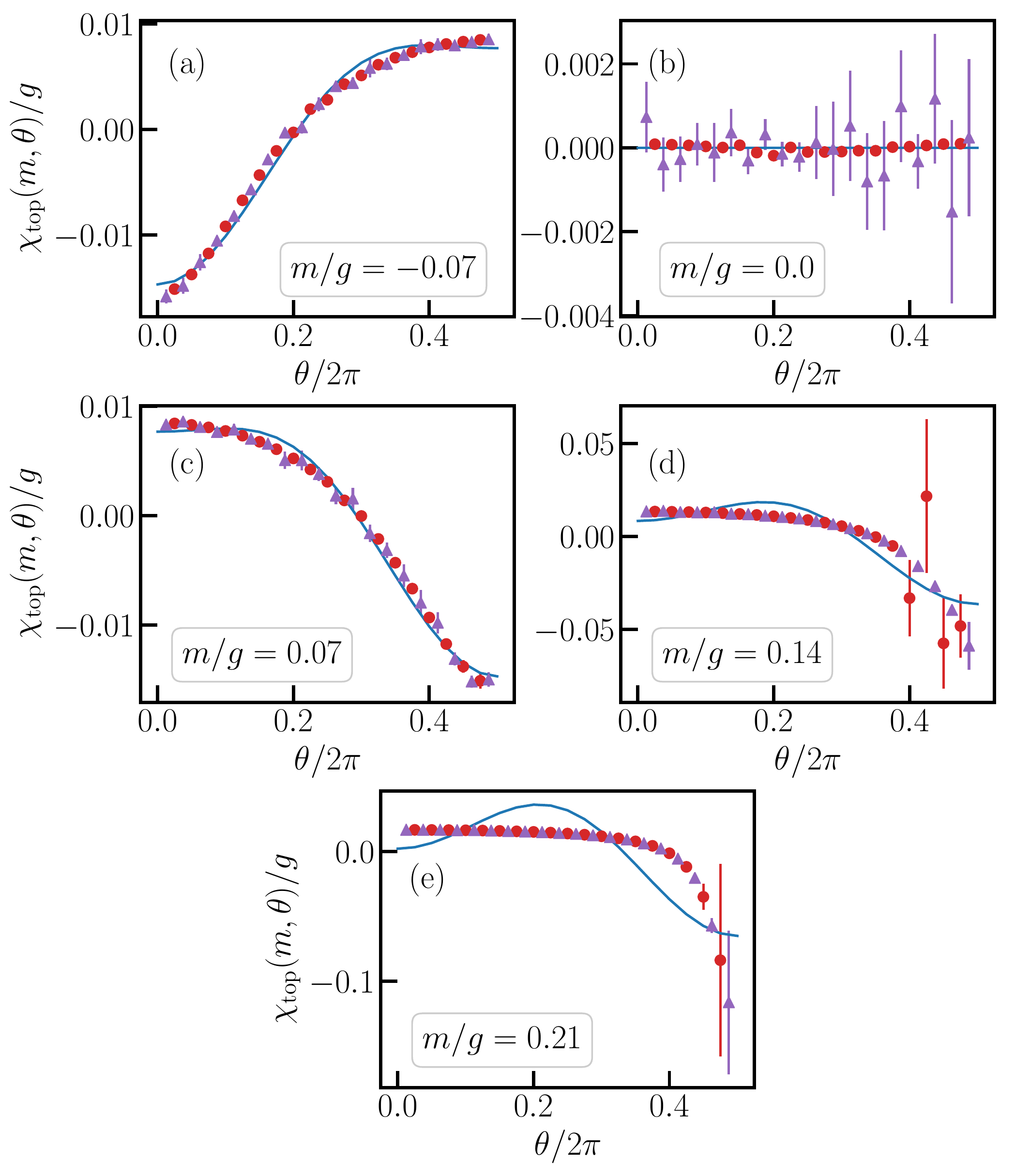}
   \caption{Topological susceptibility as a function of the angle $\theta$ for (a) $m/g=-0.07$, (b) $m/g=0.0$, (c) $m/g=0.07$, (d) $m/g=0.14$, and (e) $m/g=0.21$. The purple triangles (red dots) correspond to the (second) derivative of the continuum data for the electric field (energy density), the blue solid line corresponds to the perturbative prediction from Eq.~\eqref{eq:susceptibility}.}
   \label{fig:susceptibility_vs_theta}
\end{figure}

Although we cannot directly measure the topological vacuum susceptibility in our Hamiltonian framework, we can obtain this quantity by either numerically differentiating our continuum estimates for the electric field or by computing the second derivative from our data for the ground-state energy (see Eq.~\eqref{eq:susceptibility} and Appendix~\ref{app:extrapolation} for details). Figure~\ref{fig:susceptibility_vs_theta} shows our results for both approaches. 

In general, both methods give remarkably consistent results. The data obtained from the second derivative of the energy density have noticeably smaller error bars, except for $m/g\geq 0.14$ and $\theta\approx \pi$. Already the electric field and the ground-state energy density showed that the perturbative result breaks down for our largest two values of $m/g$, and we see that in the topological vacuum susceptibility as well, as expected (see Figs.~\ref{fig:susceptibility_vs_theta}(d)--(e)).

\begin{figure}
   \centering   
   \includegraphics[width=0.48\textwidth]{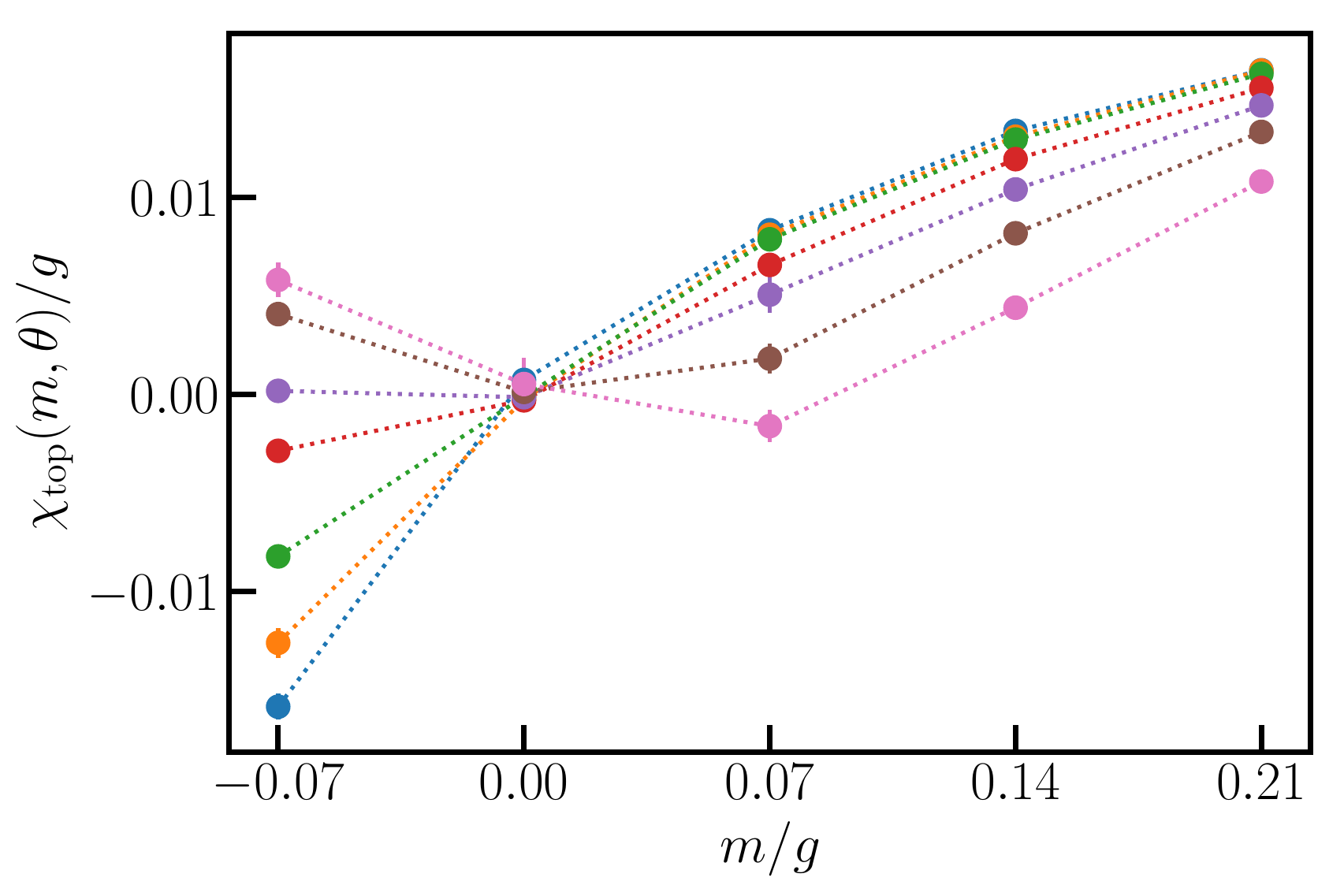}
   \caption{Topological susceptibility as a function of the bare fermion mass $m/g$ for various values of $\theta/2\pi$, namely $0.0$ (blue), $0.05$ (orange), $0.1$ (green), $0.15$ (red), $0.2$ (purple), $0.25$ (brown), and $0.3$ (pink markers). The results were obtained from numerically computing the derivative of the electric field. As a guide for the eye the markers are connected with dotted lines.}
   \label{fig:susceptibility_vs_mass}
\end{figure}

For small masses, our data is again in excellent agreement with the predictions from mass-perturbation theory, as Figs.~\ref{fig:susceptibility_vs_theta}(a)--(c) reveal. In particular, Fig.~\ref{fig:susceptibility_vs_theta}(b) shows that for vanishing fermion mass our data are compatible with $\chi_\text{top}/g = 0$, once more indicating that the background field gets completely screened in that case and $\theta$ is not a physical parameter. This becomes even more apparent in Fig.~\ref{fig:susceptibility_vs_mass}, where we show our data for the susceptibility as a function of the bare fermion mass for various values of $\theta$~\footnote{Note that we chose first-order finite differences here in order to obtain the value of the topological vacuum susceptibility at $\theta =0$ where it is commonly defined at. For the second-order finite differences, as used in Fig.\ \ref{fig:susceptibility_vs_theta}, these $\theta$-values correspond to $\theta/2\pi\rightarrow \theta/2\pi + 0.0125$.}. The figure clearly shows that for vanishing bare fermion mass, the different curves intersect at $\chi_\text{top}/g \approx 0$, which demonstrates the CP invariance of the Schwinger model for $m/g=0$. Just as in QCD, where the topological vacuum susceptibility is a measure of CP violation, the presence of a massless fermion allows the $\theta$-angle to be rotated away by an axial fermion rotation, thus CP is preserved. The same rotation also maps our results for negative and positive masses when shifting $\theta\rightarrow\theta+\pi$ (see Figs.~\ref{fig:susceptibility_vs_theta}(a) and \ref{fig:susceptibility_vs_theta}(c)), as we already observed for the ground-state energy and the electric field. Note that Fig.~\ref{fig:susceptibility_vs_mass} does not reveal this mapping, since the susceptibility is only shown for small values of $\theta\ll\pi$.

Finally, we point out that the topological susceptibility at $\theta =0$ (corresponding to the blue markers in Fig.~\ref{fig:susceptibility_vs_mass}) becomes negative for negative fermion masses and is expected to diverge to negative infinity at $m/g\approx -0.33$ (see Ref.~\cite{Creutz:2013xfa} for a discussion of a similar effect in two-flavor QCD). This is because the well-known phase transition at $m/g\approx 0.33$ and $\theta =\pi$ is equivalent to a phase transition at $m/g\approx -0.33$ and $\theta =0$, due to the above-mentioned mapping. Thus, the diverging susceptibility at $\theta =0$ is associated with the diverging correlation length as one approaches the critical point. This phase transition would be non-trivial to study in the two-flavor Schwinger model with two masses of opposite sign, giving rise to the CP-violating Dashen phase~\cite{Dashen1970}.

\section{Conclusion}

In this paper, we systematically explore the topological vacuum structure of the Schwinger model with a $\theta$-term. In particular, we study the $\theta$-dependence of the ground-state energy density, the electric field, the chiral condensate, and the topological vacuum susceptibility at zero and negative fermion mass. This mass regime is especially interesting for models with a sign problem, which require the development and testing of new numerical techniques beyond the conventional MCMC approach. The prime example would be (3+1)-dimensional QCD, where the zero-mass case was proposed as a possible solution to the strong CP problem. 

Addressing the Hamiltonian lattice formulation of the Schwinger model with numerical methods based on MPS, we show that we can reliably compute the ground state of the model in a controlled manner with small errors. While at very small masses we find excellent agreement with mass-perturbation theory, thus scrutinizing our approach, we also demonstrate the limitations of perturbative methods.

Our results provide us with a comprehensive picture of the topological vacuum structure of QED in 1+1 dimensions. For small masses, the $\theta$-dependence follows the perturbative analytical calculation from Ref.~\cite{Adam1997}. In the chiral limit, our data confirm that the model becomes CP invariant as the topological vacuum susceptibility vanishes and the $\theta$-dependent electric background field gets screened due to vacuum polarization. Thus, for $m/g=0$, the $\theta$-parameter that labels the topologically non-trivial vacua of the Schwinger model becomes an unphysical parameter, just as in the (3+1)-dimensional analog of QCD with a massless up quark~\cite{Georgi1981,Kaplan1986,Choi1988,Banks1994} or with an axion~\cite{Peccei1977,Weinberg1977,Wilczek1978}. As we go to larger masses, the perturbative prediction eventually breaks down and especially the chiral condensate deviates significantly from it.

Comparing our finite-lattice and continuum data, we find that lattice artifacts reintroduce the $\theta$-dependence of the observables in the massless limit. Moreover, in the massive regime, lattice artifacts enter inversely and with different strengths for opposite mass sign, which renders the negative-mass regime non-trivial on the lattice. In the continuum, negative masses can be trivially mapped to positive masses by shifting $\theta\rightarrow \theta +\pi$ due to the quantum anomaly~\cite{Adler1969,Bell1969,Adam1993}, which gets confirmed by our data. Our results demonstrate that MPS work well even for negative fermion masses, which has not been explored before and becomes particularly relevant in the many-flavor case, where a negative mass can generate a second-order phase transition to the CP-violating Dashen phase~\cite{Dashen1970}. 

Regarding the topological vacuum angle $\theta$, there are several interesting aspects that could be studied in the future. Generalizing our MPS setup for the Schwinger model to multiple flavors would be straightforward~\cite{Banuls2016a,Banuls2016b}, allowing us to study the above-mentioned Dashen phase~\cite{Dashen1970}. Moreover, our MPS approach is not limited to the Abelian case~\cite{Kuehn2015,Sala2018,Sala2018a,Banuls2018a,Silvi2016,Silvi2019} and could offer the possibility to explore non-Abelian gauge models in the presence of a topological $\theta$-term.

\begin{acknowledgments}
This research was supported in part by Perimeter Institute for Theoretical Physics. Research at Perimeter Institute is supported by the Government of Canada through the Department of Innovation, Science and Economic Development Canada and by the Province of Ontario through the Ministry of Economic Development, Job Creation and Trade.

Computations were made on the supercomputer Mammouth Parallèle 2 from University of Sherbrooke, managed by Calcul Québec and Compute Canada. The operation of this supercomputer is funded by the Canada Foundation for Innovation (CFI), the ministère de l'Économie, de la science et de l'innovation du Québec (MESI) and the Fonds de recherche du Québec - Nature et technologies (FRQ-NT). 
\end{acknowledgments}

\appendix

\section{Origin of $\theta$-term in continuum formulation of Schwinger Hamiltonian\label{app:theta_mass}}

In this appendix, we explain the origin of the $\theta$-term in the Hamiltonian density of the continuum Schwinger model~\eqref{eq:continuum_hamiltonian}. On the classical level, the $\theta$-term can be stripped away when the Hamiltonian is formulated in terms of the electric field~\cite{Jackiw:1983nv,Tong2018}. However, Coleman argued on physical grounds that the $\theta$-term can be understood as a background electric field in the quantized theory~\cite{Coleman1976}. A rigorous derivation was done in Ref.~\cite{Manton1985}, which also showed there are several technical subtleties. To make the paper self-contained, we here present this computation, which gets simplified in the bosonized formalism and addresses all technical challenges arising in the Hamiltonian formulation. Extensions to higher dimensions would be possible following Ref.~\cite{Dunne:1989gp}.

The bosozined version of the continuum Schwinger Hamiltonian in Eq.~\eqref{eq:continuum_hamiltonian} reads
\begin{align}
\begin{split}
\cal{H}=\,&\frac{1}{2}m^2\Phi(0)^2+\frac{1}{2}\Pi(0)^2\\&+\frac{1}{2}\sum_{p\neq0}[\Pi^+(p)\Pi(p)+(p^2+m^2)\Phi^+(p)\Phi(p)]\\
=&-\frac{g^2}{4\pi}\frac{d^2}{dA_x^2}+V(A_x)\\&+\frac{1}{2}\sum_{p\neq0}[\Pi^+(p)\Pi(p)+(p^2+m^2)\Phi^+(p)\Phi(p)]
\label{eq:Hb}
\end{split}
\end{align}
plus an irrelevant constant. Here, $m=g/\sqrt{\pi}$ is the Schwinger mass and $\Phi(p)$ and $\Pi(p)$ are the bosonic operators that satisfy canonical commutation relations and hermeticity properties. In the following subsections, we will discuss the properties of these bosonic operators for $p\neq 0$ and $p=0$ and compute the (non)conservation laws of the corresponding bosonic currents. Finally, we will use the anomaly equation~\eqref{eq:anomaly} obtained with the bosonized Hamiltonian~\eqref{eq:Hb} to demonstrate that a constant shift in the electric field of the fermionic Schwinger Hamiltonian is equivalent to an axial rotation of the massive fermionic field.

\subsection{Bosonic operators for non-zero momentum}

The bosonic operators $\Phi(p)$ and $\Pi(p)$ of the Hamiltonian density~\eqref{eq:Hb} are defined for $p\neq 0$ as
\begin{align}
\Phi(p)&=-\frac{1}{\sqrt{2}ip}[\rho_1(p)+\rho_2(p)]\label{eq:Phip}\\
\Pi(p)&=\frac{1}{\sqrt{2}}[\rho_1(p)-\rho_2(p)],
\end{align}
in terms of the bosonic chiral-charge density operators
\begin{equation}
\rho_\alpha(p)=\sum_k a^+_{\alpha,k+p}a_{\alpha,k}.
\end{equation}
Here, the index $\alpha=1,2$ denotes the left-handed ($\alpha=1$) and right-handed ($\alpha=2$) states, $p,k\in \mathds{Z}$ are the integer momenta, and the momentum-space Fermi operators $a_{\alpha,k}$ are related to the real-space Fermi operators $\psi_\alpha(x)$ via
\begin{equation}
\begin{pmatrix}\psi_1(x)\\\psi_2(x)\end{pmatrix}=\frac{1}{\sqrt{2\pi}}\sum_k\begin{pmatrix}a_{1,k}\\a_{2,k}\end{pmatrix}e^{ikx}.
\end{equation}
The bosonic currents for $p\neq 0$ are given by
\begin{align}
j^0(p)&=j_5^1(p)=\sqrt{2}ip\Phi(p)\label{eq:cPhi}\\
j^1(p)&=j_5^0(p)=\sqrt{2}\Pi(p).\label{eq:cPi}
\end{align}
In the Coulomb gauge, $\partial_x A_x=0$, the bosonic operator $\Phi(p)$~\eqref{eq:Phip} can be expressed in terms of the longitudinal part of the electric field,
\begin{align}
\Phi(p)&=-\frac{1}{\sqrt{2}g^2}{\cal{F}}_x^{\rm long}(p),\label{eq:Elong}
\end{align}
by using Gau{\ss}'s law,
\begin{align}
-\partial_x\partial_x A_t&=g^2 \psi^+\psi=g^2 j^0,\label{eq:Gauss}
\end{align}
and applying the relations $\partial_x=-ip$ and $\partial_x A_t=-{\cal{F}}_x^{\rm long}(p)$ to the left-hand side of Eq.~\eqref{eq:Gauss} and the current-operator relation~\eqref{eq:cPhi} to the right-hand side. Since the total electric charge is zero, the longitudinal part of the electric field ${\cal{F}}_x^{\rm long}(p)$ has no $p=0$ component, which is the quantization constraint in the Coulomb gauge.

\subsection{Bosonic operators for zero momentum}

The zero-momentum scalar field operators $\Phi(0)$ and $\Pi(0)$ read
\begin{align}
\Phi(0)&=\frac{i}{\sqrt{2}}\frac{d}{d A_x}=-\frac{1}{\sqrt{2}g^2}{\cal{F}}_x^{\rm tr}(0)\label{eq:Etr}\\
\Pi(0)&=\sqrt{2}V(A_x),
\end{align}
where $V(A_x)$ is the potential and ${\cal{F}}_x^{\rm tr}(0)$ is the transverse part of the electric field
\begin{equation}
{\cal{F}}_x^{\rm tr}=\frac{-ig^2}{2\pi}\frac{d}{dA_x},\label{eq:Etr2}
\end{equation}
which is classically given by $\partial_t A_x$. The transverse part of the electric field \eqref{eq:Etr2} only contains the constant $p=0$ component, i.e., ${\cal{F}}_x^{\rm tr}(0)=2\pi {\cal{F}}_x^{\rm tr}$.

\subsection{(Non)conservation laws of bosonic currents}

The Hamiltonian density~\eqref{eq:Hb} yields the following equations of motion for the operators $\Phi(p)$ and $\Pi(p)$:
\begin{align}
\frac{d}{dt}\Phi(p) &= \frac{\partial H(p)}{\partial \Pi(p)}= \Pi(p)\label{eq:eomPhi}\\
\frac{d}{dt}\Pi(p) &= -\frac{\partial H(p)}{\partial \Phi(p)}=(-p^2-m^2)\Phi(p).\label{eq:eomPi}
\end{align}
From Eqs.~\eqref{eq:cPhi}, \eqref{eq:cPi}, and \eqref{eq:eomPhi} we obtain 
\begin{equation}
\frac{d}{dt}\Phi(p)- \Pi(p) = \frac{1}{\sqrt{2}ip}\left(\frac{d}{dt}j^0(p)+\frac{d}{dx}j^1(p)\right)=0,
\end{equation}
where we used $d/dx=-ip$. This is the well-known conservation law for the vector current, 
\begin{equation}
\partial_\mu j^\mu=0.
\end{equation}
From Eqs.~\eqref{eq:cPhi}--\eqref{eq:Elong}, \eqref{eq:Etr}, and \eqref{eq:eomPi} we find
\begin{align}
\frac{d}{dt}\Pi(p) + p^2\Phi(p)&=-m^2\Phi(p)\label{eq:m2}\\
\Leftrightarrow \frac{1}{\sqrt{2}}\left(\frac{d}{dt}j_5^0(p)+\frac{d}{dx}j_5^1(p)\right)&=-\frac{g^2}{\pi}\left(-\frac{1}{\sqrt{2}g^2}{\cal{F}}_x\right),\nonumber
\end{align}
where we used $m=g\sqrt{\pi}$ and ${\cal{F}}_x={\cal{F}}_x^{\rm long}(p)+{\cal{F}}_x^{\rm tr}(0)$. This is the well-known quantum anomaly equation for the axial current,
\begin{equation}
\partial_\mu j_5^\mu={\cal{F}}_x/\pi.\label{eq:anomaly}
\end{equation}
We note that $p^2\Phi(0)=-d^2\Phi(0)/dx^2$ vanishes on the left-hand side of Eq.~\eqref{eq:m2}, therefore the zero-momentum contribution of $\Phi(p)$ only shows up in the non-derivative term on the right-hand side. For the non-derivative term, we have to add up both the $p\neq 0$ and the $p=0$ contributions given by Eqs.~\eqref{eq:Elong} and \eqref{eq:Etr}.

\subsection{Axial rotation of fermionic fields}

Using the quantum anomaly equation~\eqref{eq:anomaly} obtained with the bosonized Hamiltonian, one can now consider the fermionic Hamiltonian~\eqref{eq:continuum_hamiltonian} to demonstrate that an axial rotation of the massive fermionic field $\psi$,
\begin{align}
\psi&\rightarrow e^{i\theta\gamma_5/2}\psi
,\label{eq:rot}
\end{align}
induces a constant shift in the electric field ${\cal{F}}_x$. The rotation~\eqref{eq:rot} by an infinitesimal angle $\theta\ll 1$ shifts the fermion mass term by
\begin{equation}
m\bar{\psi}\psi\rightarrow im\theta\bar{\psi}\gamma_5\psi
\end{equation}
and analogously the chiral fermion condensate by
\begin{equation}
\langle\bar{\psi}\psi\rangle\rightarrow i\theta\langle \bar{\psi}\gamma_5\psi\rangle.
\end{equation}
Since the fermion mass perturbatively corrects the divergence~\eqref{eq:anomaly} of the axial current $j_5^\mu=\bar{\psi}\gamma_5\gamma^\mu\psi$ by
\begin{equation}
\partial_\mu j_5^\mu=-2im\bar{\psi}\gamma_5\psi+\frac{{\cal{F}}_x}{\pi},\label{eq:anomaly2}
\end{equation}
the axial rotation of the fermionic field~\eqref{eq:rot} induces the following shift in the Hamiltonian density:
\begin{align}
\mathcal{H} \rightarrow \mathcal{H} + \frac{\theta}{2} \partial_\mu j_5^\mu &=\mathcal{H}+i\theta m\bar{\psi}\gamma_5 \psi + \theta \frac{{\cal{F}}_x}{2\pi},
\end{align}
where we used $\gamma_5\gamma^0=-\gamma^0\gamma_5$. Thus, the angular parameter $\theta$ in the fermion mass term can be absorbed by a shift in the electric field, ${\cal{F}}_x\rightarrow {\cal{F}}_x+\theta/2\pi$, and vice versa. In particular, for $m=0$ the $\theta$-parameter becomes unphysical, since it can be rotated away without affecting any mass term, which corresponds to absorbing the phase of the chiral condensate $\langle\bar{\psi}\psi\rangle$ in the Schwinger boson field.

\section{Details of the extrapolation procedure\label{app:extrapolation}}

Here we give some details how we control the errors in our numerical simulations and how we extrapolate our data to obtain first the thermodynamic limit and finally the continuum limit. Since the extrapolation procedure has to be done independently for each combination of $(m/g,\theta)$, we suppress all arguments in the following and just refer to $E_0/2N$, $F_\text{av}$, and $C$, meaning that we look at these quantities at a specific value of $(m/g,\theta)$.

\begin{figure}
   \centering
   \includegraphics[width=0.48\textwidth]{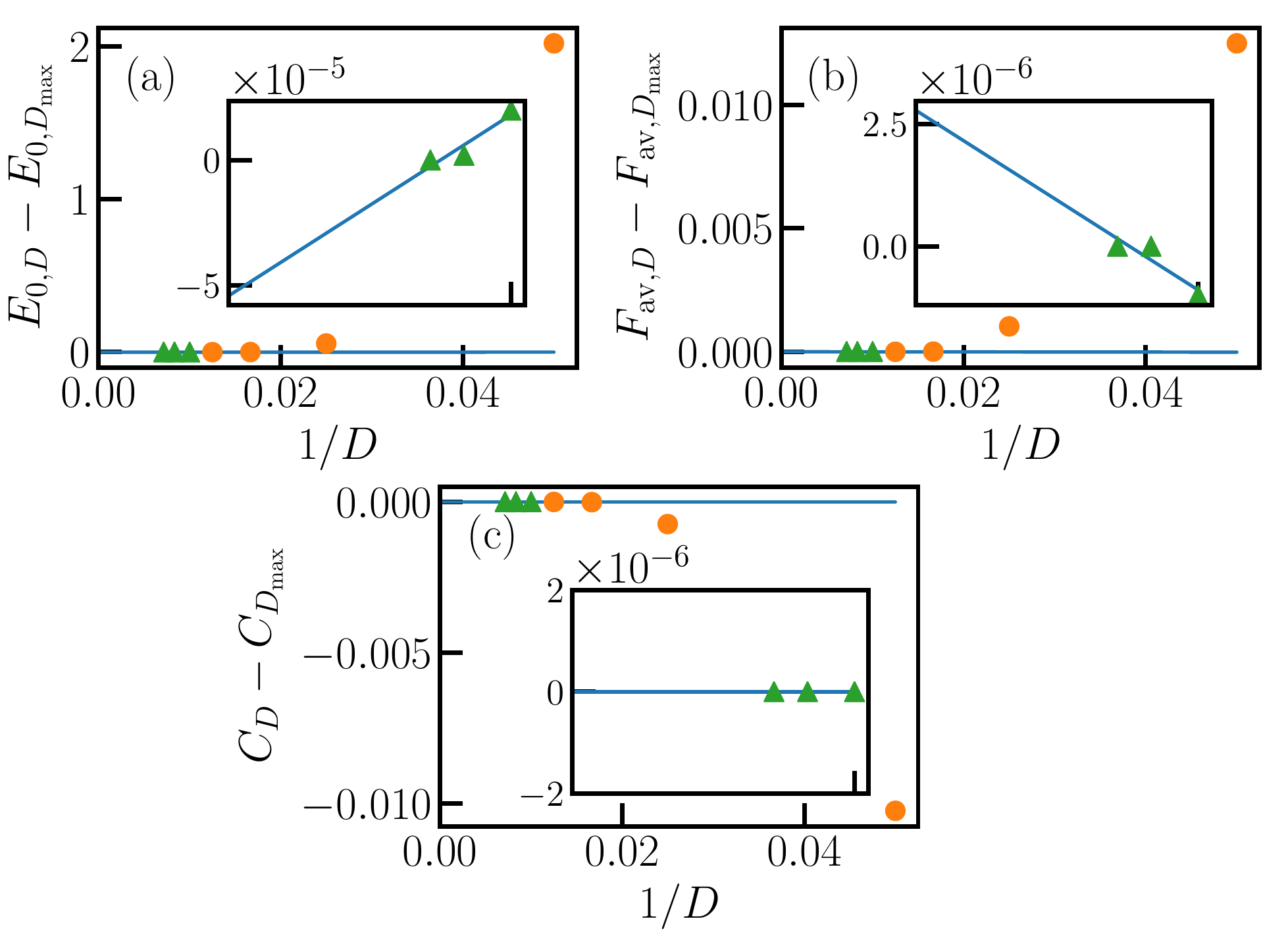}
   \caption{Extrapolation to infinite bond dimension, $1/D\rightarrow 0$, for (a) the ground-state energy, (b) the electric field, and (c) the chiral condensate, for exemplary values of $x=160$, $m/g=0.07$, $N=354$, and $\theta=0.2$. In all panels, the blue solid line represents a linear fit to the values for the largest three bond dimensions (depicted as green triangles). The insets show the regions around the data used for the extrapolation in greater detail.}
   \label{fig:extrapolation_bond_dimension}
\end{figure}

In a first step, for every combination of $(N,x,m/g,\theta)$, we estimate the numerical error due to the finite matrix size in our MPS ansatz. To this end, we plot the observables $O$ that we measure as a function of $1/D$, and we use the three data points with the largest values for $D$ to linearly extrapolate to the limit $D\to\infty$ (see Fig.~\ref{fig:extrapolation_bond_dimension} for
an example). We proceed in a standard manner: the central value is taken to be the mean value of our data point with the largest bond dimension, $O_{D_\text{max}}$, and the extrapolated value $O_{D = \infty}$. The error is estimated as half of the absolute value of their difference, $\delta O_D = |O_{D_\text{max}} - O_{D = \infty}|/2$. In general, we find that our bond dimensions are large enough to avoid noticeable truncation effects due to the finite matrix size. In addition, we have another error due to the finite convergence tolerance $\eta$ in our simulations which results in $\delta E_{0,\eta} = \eta E_{0,D_\text{max}}$ for the ground-state energy and $\delta O_{\eta} = \sqrt{\eta} O_{D_\text{max}}$ for other local observables~\cite{Haegeman2011a}. The total error is then estimated as the square root of the sum of squares, $\delta O = \sqrt{(\delta O_D)^2 + (\delta O_{\eta})^2}$.

\begin{figure}
   \centering
   \includegraphics[width=0.45\textwidth]{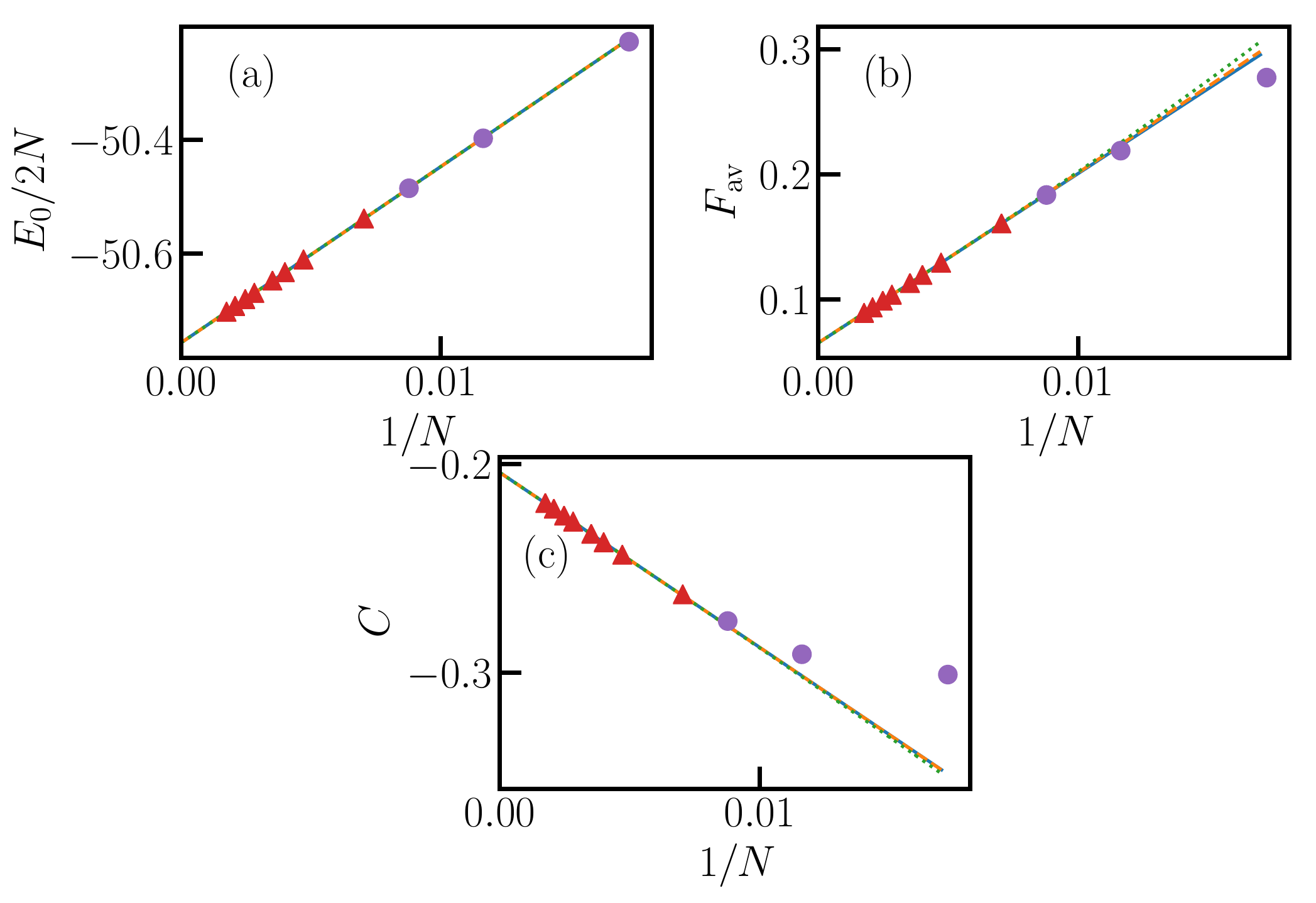}
   \caption{Extrapolation to infinite system size, $1/N\rightarrow 0$, for (a) the ground-state energy, (b) the electric field, and (c) the chiral condensate, for exemplary values of $x=160$, $m/g=0.07$, and $\theta=0.2$. In all panels the red triangles correspond to the data used to extrapolate to the thermodynamic limit, the blue solid lines to the best linear fit, the orange dashed lines to the best quadratic fit, and the green dotted lines to the best cubic fit in $1/N$.}
   \label{fig:extrapolation_N}
\end{figure}

After estimating the numerical errors due to the finite matrix size, we extrapolate our data for each combination of $(x,m/g,\theta)$ to the infinite-volume limit $N\to\infty$, where we propagate our errors from the extrapolation in $D$. In general, we observe strong finite-volume effects for volumes $N/\sqrt{x} < 15$, thus for the extrapolation we only consider volumes larger than that. To estimate the infinite-volume limit, we fit our data to polynomials in $1/N$ up to degree $3$ (see Fig.~\ref{fig:extrapolation_N} for an example). For each polynomial, we try every fitting interval of consecutive data points, which contains at least two more data points than the degree of the polynomial. To obtain the central value, we choose the fit with the lowest value of $\chi^2_\mathrm{d.o.f.}$. In case we have several fits with $\chi^2_\mathrm{d.o.f.}<1$, we choose the polynomial of smallest degree in $1/N$ that achieves this value. In most cases, we find that a linear fit in $1/N$ is enough to describe our data well. In addition to the error of the fitting coefficient, we estimate our systematic error. To this end, we compare our central value to the value obtained
from the next best fit using the same degree polynomial or to the one obtained with the next highest order. The total error is then estimated, analogously to the extrapolation in $D$, as the square root of the sum of squares.

\begin{figure}
\centering
\includegraphics[width=0.45\textwidth]{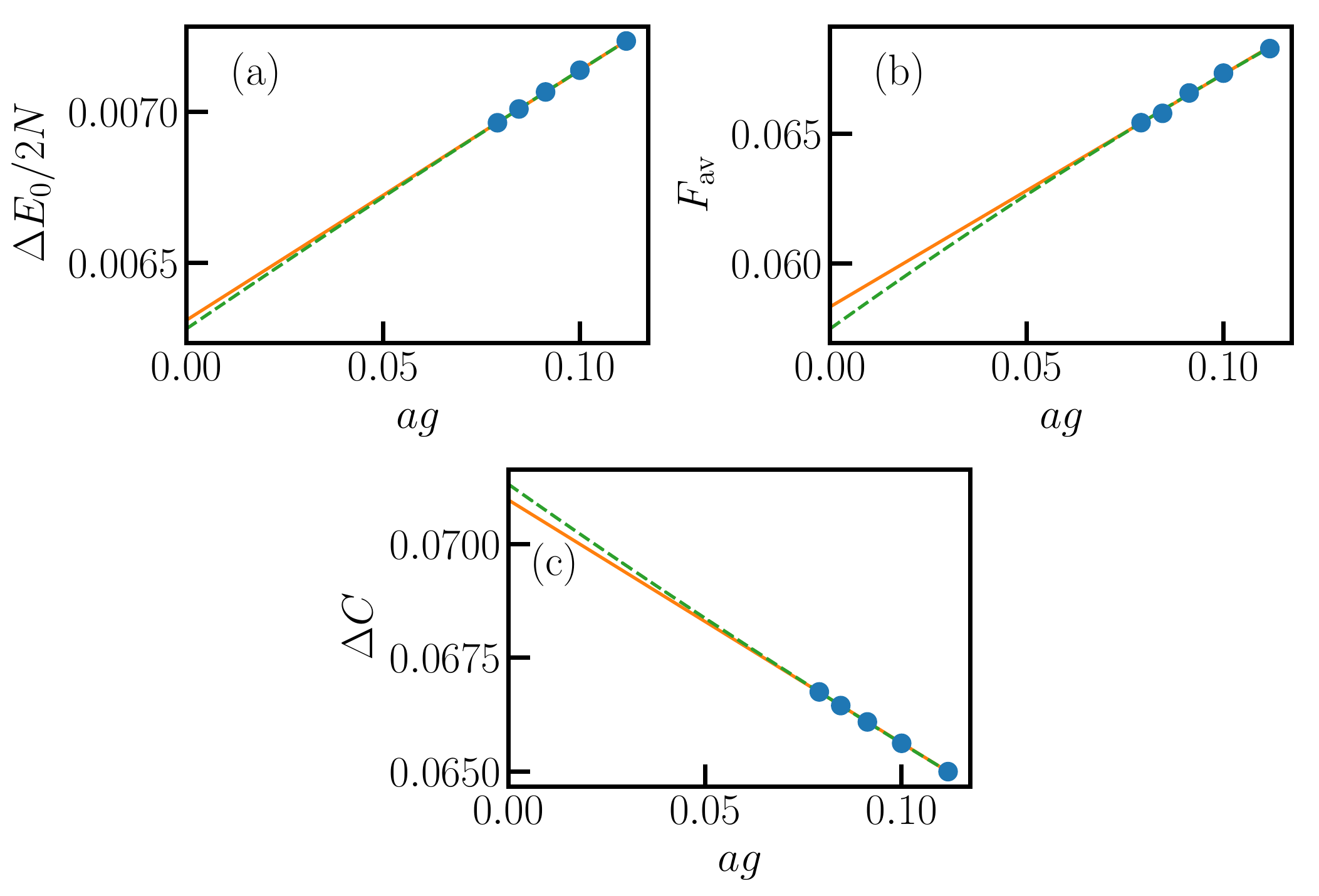}
\caption{Extrapolation to the continuum, $ag\rightarrow 0$, for (a) the ground-state energy, (b) the electric field and (c) the chiral condensate for exemplary values of $m/g=0.07$, and $\theta=0.2$. The blue dots show the data obtained after extrapolating to the thermodynamic limit, the solid line a linear fit, and the dashed line a quadratic fit in $ag$.}
\label{fig:extrapolation_x}
\end{figure}

In a final step, we extrapolate our data to the continuum corresponding to $ag\to 0$. To this end, we fit our finite-lattice data again to polynomials in $ag$. In general, we observe that a linear function or sometimes even a constant is enough to describe our data well (see Fig.~\ref{fig:extrapolation_x} for an example). We again propagate the errors from the extrapolation in $N$ to estimate the final error of our data.

To obtain the electric field from our continuum estimates for the energy density, we follow Eq.~\eqref{eq:field_density} and compute the derivative numerically using second-order finite differences~\footnote{Note that we use
\begin{align*}
   f'(x) = \frac{1}{2h}\bigl(f(x+h) - f(x-h)\bigr) +\mathcal{O}(h^2)
\end{align*}
with $x = \theta + \Delta \theta / 2$ and $h = \Delta \theta / 2$.
 }
\begin{align}
   \begin{aligned}
   &\frac{\mathcal{F}(m,\theta +\Delta \theta/2)}{g} \approx\\
    &\quad\quad2\pi\frac{\Delta {\cal{E}}_0(m,\theta + \Delta \theta)/g^2 -\Delta {\cal{E}}_0(m,\theta)/g^2}{\Delta \theta}.
   \end{aligned}
\end{align}
For all the data we show, the distance between two different points is $\Delta \theta = 0.025$. The error of the electric field is then estimated by propagating the error in the UV-finite energy density as a systematic error.

The topological vacuum susceptibility can be calculated in a similar fashion, we can either numerically differentiate our results for the UV-finite energy density twice with respect to $\theta$,
\begin{align}
   \begin{aligned}
      \frac{\chi_\text{top}(m,\theta)}{g} \approx -\frac{1}{\Delta \theta^2}&\left(\frac{\Delta {\cal{E}}_0(m,\theta + \Delta \theta)}{g^2} - 2\frac{\Delta {\cal{E}}_0(m,\theta)}{g^2} \right.\\
      &\left. \phantom{++}+ \frac{\Delta {\cal{E}}_0(m,\theta - \Delta \theta)}{g^2}\right),
   \end{aligned}
\end{align}
or compute the derivative of our results for the electric
field,
\begin{align}
   \frac{\chi_\text{top}(m,\theta + \Delta \theta/2)}{g} \approx \frac{\mathcal{F}(m,\theta + \Delta \theta) -  \mathcal{F}(m,\theta )}{2\pi\Delta \theta},
\end{align}
where again the distance between two different angles is $\Delta \theta = 0.025$ and we propagate the errors in the UV-finite energy density and in the electric field values as systematic errors to obtain an error estimate for the topological susceptibility.

\section{Data for a full period of $\theta$\label{app:fullperiod}}

\begin{figure}[H]
   \centering
   \includegraphics[width=0.48\textwidth]{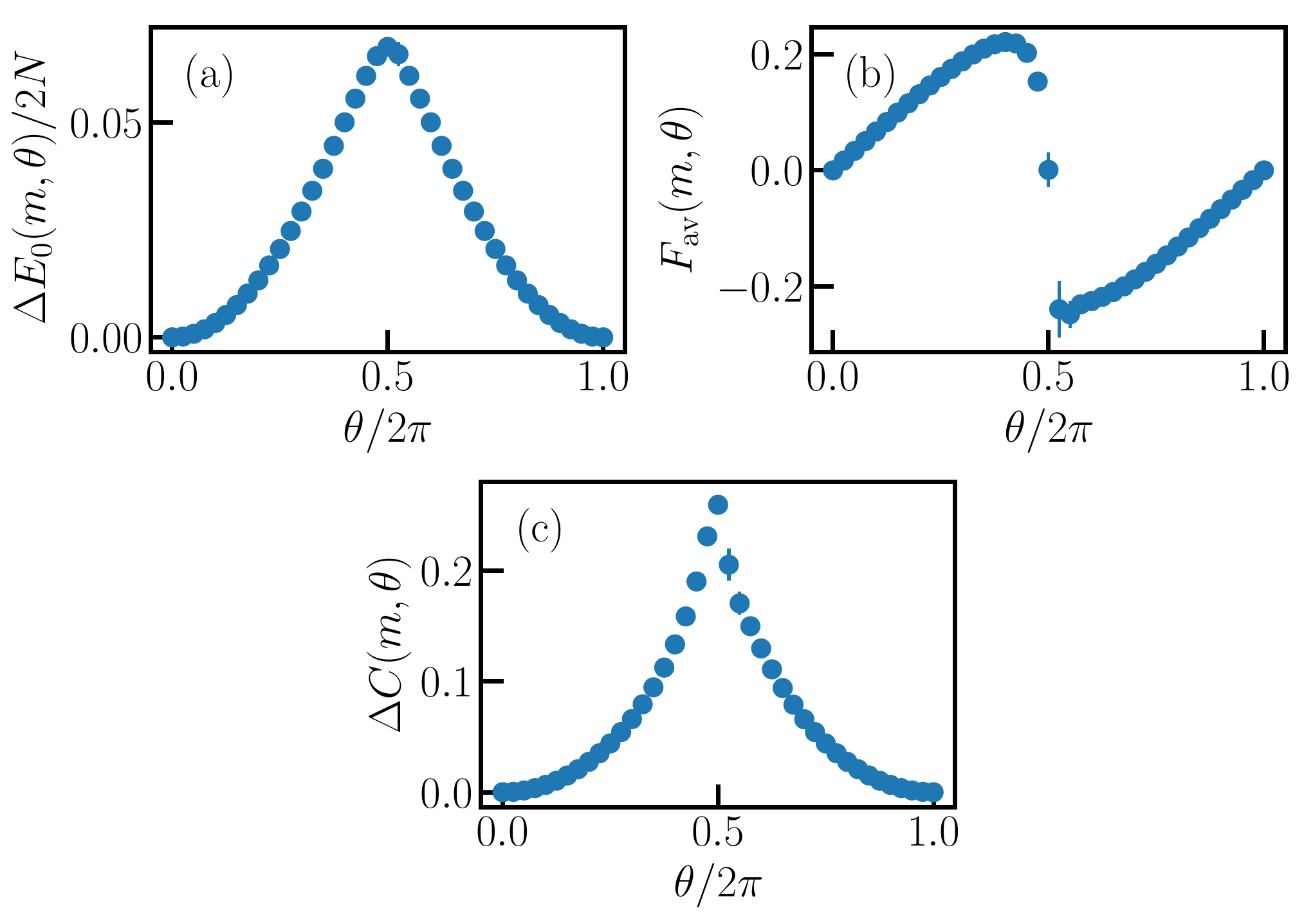}
   \caption{(a) Energy density, (b) electric field, and (c) chiral condensate after extrapolating to the thermodynamic limit over a full period for $m/g=0.21$ and $x=80$.}
   \label{fig:full_cycle_mg021}
\end{figure}

In the main text, we focused on the regime $0\leq\theta\leq \pi$, since all quantities studied are (point) symmetric around $\pi$. Figure~\ref{fig:full_cycle_mg021} shows an explicit example of a full period for $m/g=0.21$ after extrapolating to the thermodynamic limit. We restrict ourselves to a single lattice spacing corresponding to $x=80$, since this value is already very close to the continuum limit for such large masses, as demonstrated by the data in the main text. Looking at Fig.~\ref{fig:full_cycle_mg021}, we see that the data for the ground-state energy density, the electric field, and the chiral condensate are indeed symmetric around $\theta=\pi$, and we can obtain precise estimates throughout the entire period of $\theta\in [0,2\pi]$.

As theoretically predicted in Ref.~\cite{Coleman1976} and numerically demonstrated in Refs.~\cite{Byrnes2002,Buyens2017}, the continuum model exhibits a first-order transition at $\theta=\pi$ for bare fermion masses larger than the critical value $(m/g)_c\approx0.33$. This is accompanied by a spontaneous breaking of the CP symmetry, and the critical line ends in a second-order quantum phase transition exactly at $(m/g)_c$. Since our value for the largest bare fermion mass is still smaller
\noindent than the critical one, we do not expect a transition to happen at $\theta=\pi$. 

At first sight, our data for the chiral condensate in Fig.~\ref{fig:full_cycle_mg021}(c) give the impression that there could nevertheless be a transition, as we observe a sharp peak at that value of $\theta$. However, due to the large slope of this quantity, our resolution in $\theta$ is limited. Taking a closer look at Fig.~\ref{fig:full_cycle_mg021}(b), we see that the electric field at $\theta=\pi$ vanishes, thus indicating that the CP symmetry is not broken~\cite{Coleman1976} and there is no phase transition. Our data for the energy density (Fig.~\ref{fig:full_cycle_mg021}(a)) corroborate this picture. Although for $\theta=\pi$ there is a noticeable peak, we do not observe a cusp in the data, thus giving another indication that for $m/g=0.21$ there is no transition. Nevertheless, the features in the energy density, the electric field, and the chiral condensate hint toward an upcoming phase transition at $(m/g)_c\approx0.33$.

\bibliography{Papers}
\end{document}